\documentclass[sn-mathphys-num]{sn-jnl}% Math and Physical Sciences Numbered Reference Style
\usepackage{graphicx}%
\usepackage{subfig}%
\usepackage{multirow}%
\usepackage{amsmath,amssymb,amsfonts}%
\usepackage{amsthm}%
\usepackage{mathrsfs}%
\usepackage[title]{appendix}%
\usepackage{xcolor}%
\usepackage{textcomp}%
\usepackage{manyfoot}%
\usepackage{booktabs}%
\usepackage{algorithm}%
\usepackage{algorithmicx}%
\usepackage{algpseudocode}%
\usepackage{listings}%
\usepackage{array}
\usepackage{hyperref}%
\hypersetup{draft}%

\theoremstyle{thmstyleone}%
\theoremstyle{thmstyletwo}%

\theoremstyle{thmstylethree}%

\begin{document}

\title[Article Title]{Forecasting of Bitcoin Prices Using Hashrate Features: Wavelet and Deep Stacking approach}


\author*[1]{\fnm{Ramin} \sur{Mousa}}\email{Raminmousa@znu.ac.ir}

\author[2]{\fnm{Meysam} \sur{Afrookhteh}}\email{Afrookhtesshs@gmail.com}

\author[3]{\fnm{Hooman} \sur{Khaloo}}\email{Khaloohooman1998@gmail.com}
\author[4]{\fnm{Amir Ali } \sur{Bengari}}\email{Amirali.bengari@ut.ac.ir}
\author[5]{\fnm{Gholamreza } \sur{Heidary}}\email{heidaryreza137@gmail.com}


%\affil*[1]{\orgdiv{Mathematics}, \orgname{Shahid Bahonar}, \city{Tehran}, \country{Iran}}}

%\affil[2]{\orgdiv{Computer}, \orgname{Azad University},  \city{Tehran}, \country{Iran}}}

%\affil[3]{\orgdiv{Computer}, \orgname{Shafir University},  \city{Tehran}, \country{Iran}}}
%\affil[4]{\orgdiv{Computer}, \orgname{Zanjan university},  \city{Zanjan}, \country{Iran}}}


\abstract{Digital currencies have become popular in the last decade due to their non-dependency and decentralized nature. The price of these currencies has seen a lot of fluctuations at times, which has increased the need for prediction. As their most popular, Bitcoin(BTC) has become a research hotspot. The main challenge and trend of digital currencies, especially BTC, is price fluctuations, which require studying the basic price prediction model. This research presents a classification and regression model based on stack deep learning that uses a wavelet to remove noise to predict movements and prices of BTC at different time intervals. The proposed model based on the stacking technique uses models based on deep learning, especially neural networks and transformers, for one, seven, thirty and ninety-day forecasting. Three feature selection models, Chi2, RFE and Embedded, were also applied to the data in the pre-processing stage. The classification model achieved 63\% accuracy for predicting the next day and 64\%, 67\% and 82\% for predicting the seventh, thirty and ninety days, respectively. For daily price forecasting, the percentage error was reduced to 0.58, while the error ranged from 2.72\% to 2.85\% for seven- to ninety-day horizons. These results show that the proposed model performed better than other models in the literature.}




\maketitle

\section{Introduction}\label{sec1}

Bitcoin is one of the first decentralized digital currencies that banks and individuals do not have any involvement in its control, which has the advantage that no institution can claim its ownership. BTC is highly popular as a pioneer in digital currencies. This money was created in 2009 but became customary in 2017. The price of BTC has risen and fallen several times throughout the year. Many economic institutions are interested in predicting the price of Bitcoin. This issue is significant for existing or potential investors and for government structures. Therefore, the demand for BTC price prediction mechanism is high\cite{kaushal2016bitcoin}.

Price volatility is essential to intangible digital assets, especially digital currencies. The value of BTC is different from the value of stocks\cite{doumenis2021critical}. Different algorithms are used in the market for price prediction. However, the factors affecting Bitcoin are different, so it is necessary to predict the value of BTC in order to make the right investment decisions. Unlike the stock market, gold, oil, etc., the price of BTC does not depend on business events or the intervention of any government. Therefore, it becomes necessary to use machine learning and deep learning technology to predict its price value. Business with machine learning (ML) and artificial intelligence (AI) has gained attention and attention in the past few years.

Much research has been done to predict the changes in BTC. Most researchers use the historical information of BTC to predict its future price using artificial neural networks (ANN) and ML methods\cite{chen2023analysis}\cite{zhu2023price}\cite{siddharth2023cryptocurrency}. Another group of methods uses sentiment analysis and makes predictions using positive or negative comments\cite{gb2023cryptocurrency}\cite{sharma2023price}. Several researchers have also identified influential people in the social network and use their opinions as a criterion for prediction\cite{dhandecryptocurrency}\cite{mittal2019short}.

In this article, two issues are considered for BTC. In the first problem, classification aims to increase and decrease based on price. For this purpose, the price of BTC was written into two classes, 0 and 1, based on the following relationship:

\begin{flalign}
Class=\begin{cases}
    1, & if\quad Next_{price}>Current_{price}\\
   0, & if\quad Next_{price}<=Current_{price}
  \end{cases}
\end{flalign}

In the second problem, the real price of BTC was considered the forecast value, and in this case, the problem was considered a regression problem. Considering these two issues, a model based on deep neural networks and stacking technique was considered. Different models make predictions based on input data after training. This prediction is used in a higher-level layer as a selection based on the majority vote in classification problems or averaging in regression problems.
Previously reviewed literature has tried to use technical data or social networks, while hash rate data is considered the proposed model's raw input. This research aims to provide a suitable approach for medium-term price forecasting and high/low forecasting for forecast horizons from 7 days to 90 days. Also, this research focuses on predicting the daily closing price and short-term prices prediction. Our results show that our stacking model are better than the latest literature on daily forecasting and price up/down classification.

\section{Related Works}\label{sec2}

Cryptocurrencies have attracted considerable awareness, and Bitcoin has been the most notable in terms of price and academic and non-academic research. Since BTC, like other digital currencies, has much volatility in its price, predicting its price is challenging. Although several researchers used traditional economic and statistical methods to discover the driving variables of Bitcoin price, the progress in developing prediction models for decision-making tools in investment techniques is still in the early stages \cite{R1}. Many cryptocurrency price forecasting methods cover objectives such as forecasting in a one-step(one-hour,one-day, or one-week) approach that can be performed through time series analysis, neural networks(NNs), and machine learning(ML) algorithms. However, it is important to understand the price trend of BTC in the long and middle term. In this section, we examine ML and DL approaches in predicting cryptocurrencies.

In \cite{R3}, the authors have turned the primary research focus to extracting the factors or features that affect the BTC price. This research used the combined prediction model with double support vector regression as the primary model. Also, the XGBoost and random forest algorithms were considered feature selection algorithms. A hybrid model with dual support vector regression (TWSVR), support vector regression (SVR) and least squares support vector regression (LSSVR) were used to predict Bitcoin price. To optimize the models, the authors used WOA and PSO algorithms. The models used six combinations: WOA-SVR, WOA-LSSVR, WOA-TWSVR, PSO-SVR, PSO-LSSVR, and PSO-TWSVR. The best-reported results combined PSO-TWSVR with RF feature selection, achieving $R^2=0.924$, MAE=0.032, MSE=0.002, RMSE=0.044, and MAPE=0.0441.

Many types of research have shown a comparison between the approaches available in the literature. In \cite{R4}, the authors have discussed the performance of two LSTM and ARIMA models for the short-term prediction of Bitcoin. For this purpose, they have two educational examples from 12/21/2020 0:00 to 12/19/2021 20:30 (using a 24-hour time format and date format is MM/DD/YYYY) and a test that starts from 12/ 19/2021 20:40 to 12/21/2021 16: were considered for cross-validation of prediction findings. The ARIMA approach reached MAE=837.77, MAPE=1.79\%, RMSE=940.40, and Accuracy=98.21\%. On the other hand, the LSTM approach reached MAE=126.97, MAPE=0.27\%, RMSE=151.95, and Accuracy=99.73\%, and the results got better. Unlike ARIMA, which could only track the Bitcoin price trend, the LSTM model could predict direction and value over time. This research demonstrates the stable capacity of LSTM to predict Bitcoin price volatility despite the complexity of ARIMA.

Advanced deep-learning approaches \cite{R5} were used to predict Ether currency. The authors used LSTM, GRU, TCN(Temporal Convolution Network), Hybrid LSTM-GRU, Hybrid LSTM-TCN, Hybrid GRU-TCN, and Ensemble that combined the predictions of the LSTM, Hybrid LSTM-GRU and Hybrid LSTM-TCN models were used for short-term and long-term forecasting. They concluded that similar factors influence the daily and weekly forecasts. These factors included EMA and MACD technical indicators, Bitcoin price and Ethereum search volume index on Google. The introduced approaches were used in both regression and classification modes for daily and weekly intervals. In daily forecasting, in the regression model, the Hybrid LSTM-GRU approach reached RMSE=8.6, and in the daily classification mode, the Ensemble approach achieved $Accuracy=84.2$. In the weekly model, the TCN regression mode obtained the best RMSE, which was 36.9. In the case of weekly classification, the Ensemble approach was still the best model, which achieved Accuracy=78.9.

Due to the high accuracy of RNN approaches in time series, the authors \cite{R6} used different approaches such as LSTM, GRU, Bi-LSTM, and Bi-GRU to predict the price of BTC. Bidirectional approaches have twice as many free parameters as standard approaches. These models receive data and its inverse as input and use both for learning in parallel. These models were considered in daily and regression models. The Bi-GRU model has achieved an RMSE value of 427.85, an MAE value of 295.43, and $R^2-score$ of 0.9986, which is better than other models. According to the results shown in this research, the models lacked any kind of overfitting, and on the other hand, the Bi-GRU model behaved similarly to the real data compared to other approaches in long-term forecasting.

There are more than 6,000 digital currencies worldwide, which shows that digital currency is a growing investment market. For this reason, low-income investors invest in promising cryptocurrencies with low market capitalization. However, these investors often invest unconsciously and lose money for various reasons. However, it is possible to make a reasonable investment using deep learning-based data analysis methods that lead to short-term or long-term profits. In \cite{R7}, a study was conducted to analyze cryptocurrencies with recurrent deep-learning methods. Five cryptocurrencies (Dollar Pax (USDP), Bitcoin Gold (BTG), OKB, Telcoin (TEL) and Audius (AUDIO)) were fed into LSTM and single-base LSTM networks. This study considered the group mode LSTM, where the recursion-sequence parameter in the first hidden layer is set to True for comparison with the single-base LSTM. This study found that LSTM network ensembles do not always perform better than single-base LSTM networks in analyzing promising cryptocurrencies. This claim was clearly seen in the Pax Dollar (USDP). In other words, these methods can be used to obtain reliable analysis results about promising cryptocurrencies.

Traditional machine learning approaches such as Random Forest, Gaussian Naive Bayes, Support Vector Machine, K-Nearest Neighbors and deep learning RNN were investigated in \cite{R81} to predict Bitcoin. This research has considered the long-term forecast of Bitcoin. They used a simple LSTM model for prediction, and this model reached an accuracy of 76.99\%, which, compared to the best-tested machine learning model, Random Forest, could achieve an improvement of about 0.15\%. Another study of Bitcoin prediction in the Yahoo Finance Stock Market using LSTM networks was conducted \cite{R8}. They proposed a simple shallow LSTM model for this purpose and showed the effect of different batch sizes for fitting inputs to LSTM and dropout in the learning process. The best result for the LSTM approach was equal to $RMSE=313.6623$, which resulted from $batch=500$ and $Dropout=0.5$. Ensemble-enabled LSTM was another approach based on the LSTM network that \ used cite{R9} to predict Bitcoin with various time intervals. In this model, three intervals were considered inputs: minute, hourly, and daily.
Moreover, an LSTM model named day model, hour model, and min model was considered for each interval. The ensemble output of this approach was obtained using the relationship $o = \alpha M + \beta H + \gamma D + \delta$, in which $\alpha, \beta$, and $\gamma$ showed the effect of each of the minute, hourly, and daily models. $\delta$ is a bias to adjust for the shift of the predicted values. Various tests were performed on the minute, hourly and daily data and Ensemble $M+H+D$ LSTM NN obtained the best result; this model achieved an accuracy of 95.56.

The authors \cite{R11} proposed a new framework called DLGueS. They evaluated its performance using the price history of similar cryptocurrencies, namely Bitcoin and Litecoin, and Twitter sentiment, to predict Dash and BTC. They used a simple model that included two variables for Twitter data related to cryptocurrency quantity and quantity information. The basic model used an LSTM+GRU network to extract features on price data, which was combined with textual data in a concatenate layer. Their proposed approach reached  $MAPE=4.7928$ on Dash and $MAPE=4.4089$ on BTC cash. In this framework, VADER used for sentiment analysis.

The success of neural networks in integrating sentiment analysis can lead to better predictions for price analysis. Another example of these studies is the article \cite{R12}. They proposed a practical multi-source emotion hybrid approach to improve performance over other evaluated approaches. Experiments were conducted using various configurations and models, from MLPs to CNN and RNN, to provide a reliable evaluation of sentiment-based DL trading strategies. In the combination of sentiment analysis and price information, the MLP approach achieved an average percentage of 224, CNN an average percentage of 228, and LSTM an average percentage of 224.

ARIMA was used in the work\cite{F2_} to forecast using the information of 7 days before and 30 days before. 7 and 30 refer to the time used as a sliding window. The information used in this research includes the historical characteristics of BTC. This model reached $R^2=0.991$, $MAPE=0.035$, $RMSE=91.2$, and $CORR=0.995$.

In \cite{F1}, the authors compared deep learning and machine learning approaches. They compared ANN, RNN, CNN, random forest (RF) and k-nearest neighbours (k-NN). According to the results, deep learning approaches recorded a longer execution time than machine approaches but were more efficient.

The authors proposed an ensemble-SVR model for data segmentation in \cite{F1_}. At first, they compared the virtual currency price change curve with other currencies. It is known that these virtual currencies have substantial price changes. When SVR is directly used to analyze data on these rapidly changing curves, it does not conform to the traditional idea that the more data, the more accurate the model and the accuracy of the prediction results are significantly affected. With the amount of data through many experiments, the authors found that the more data samples there are for a strongly changing curve, the more sparsely these samples are distributed in the edge region by SVR, resulting in lower prediction accuracy. Their proposed model for data segmentation reached $RMSE=64.59$ within 30 days, while the SVR model reached $RMSE=555.89$.

Some other researchers have mapped the descriptive interval into two up/down classes, where class 1 is an up class and zero class a down. This problem is a binary classification problem, usually called sentiment analysis. In this regard, some Twitter and historical features have been used for this purpose.

\section{NETWORK DESCRIPTION}

\begin{figure*}[!hptb]
\begin{center}
\includegraphics[width=1\textwidth]{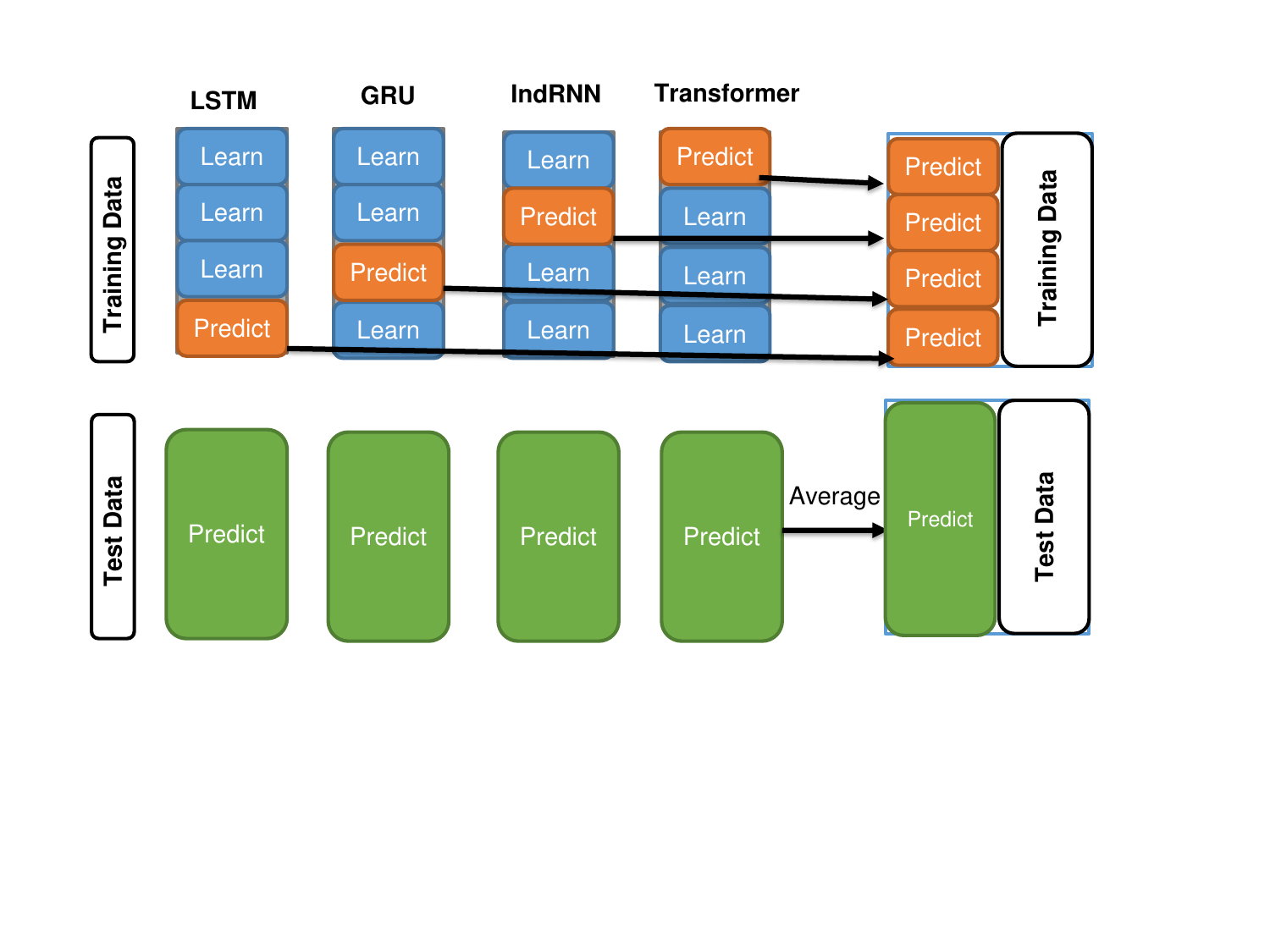}
\caption{Proposed Stack models.}
\label{CNS}
\end{center}
\end{figure*}

An overview of the proposed approach deep stacking model is given in Figure \ref{CNS}. This work presents a deep stack learning framework for time series using a prediction scheme based on cumulative and average learning to take deep and consistent features for price prediction one step further. The framework consists of two steps:

\begin{enumerate}
  \item Data preprocessing using wavelet transform, which is applied to decompose the Bitcoin price time series and to remove noise
  \item A deep Stacked model to predict overall results.
\end{enumerate}

\subsection{Wavelet transform for time series}
 This study uses Wavelet transform to remove data noise because it can handle non-stationary time series data. The key feature of the wavelet can analyze the frequency components of time series simultaneously with time compared to the Fourier transform. This study uses the Haar function as the basis wavelet function because it can decompose the time series into the time and frequency domains and significantly reduce the processing time \cite{F2}. A similar study of this series has been used in the work \cite{W1}. Continuous wavelet transform (CWT) for time series data can be defined as\cite{W1}:

\begin{equation}\label{s}
  \phi_{\gamma,\eta}(t)=\frac{1}{\sqrt{a}}\phi(\frac{t-\eta}{\gamma})
\end{equation}

 $\gamma$ is the scaling factor. $\phi_(t)$ is the basic wavelet that follows the wavelet acceptance rule according to studies \cite{W2}:

\begin{equation}\label{s}
  C_\Phi=\int_0^\infty \frac{|\Phi(\omega)|}{2}dw<\infty
\end{equation}

Where $\Phi(\omega)$ is a function of the frequency $\omega$ and also the Fourier transform of $\Phi(t)$. Let $x(t)$ denote a quadratic function $(x(t) \in L^2(R))$. Then CWT with wavelet $\Phi$ can be defined as follows for time series:

\begin{equation}\label{s}
  CWT_x(\gamma,\eta)=\frac{1}{\sqrt{\gamma}}\int_{-\infty}^{+\infty} x(t)\hat{\Phi(\frac{t-\eta}{\gamma})}dt
\end{equation}

where $\hat{\Phi(t)}$ represents its complex conjugate function. The inverse transform of the continuous wavelet transform can be shown as follows:
\begin{equation}\label{d}
  x(t)=\frac{1}{C_\phi}\int_{0}^{+\infty}\frac{d\gamma}{\gamma^2}\int_{-\infty}^{+\infty} CWT_x(\gamma, \eta)\phi_{\gamma,\eta}(t)d\eta
\end{equation}

Since the CWT coefficients have significant redundant information, it is reasonable to sample the coefficients to reduce redundancy. Mallat \cite{W3} proposed time series filtering using high-pass filter and low-pass filter to implement discrete wavelet transform. This algorithm has two wavelet types: father wavelet $\xi(t)$ and mother wavelet  $\varsigma(t)$. The  $\xi(t)$ and $\varsigma(t)$ are integrated with 1 and 0, respectively, which can be formulated as follows:
\begin{equation}\label{g}
  \xi_{m,n}(t)=2^{-\frac{1}{2}}\xi(2^{-j}-k)
\end{equation}
\begin{equation}\label{s}
  \varsigma_{m,n}(t)=2^{-\frac{1}{2}}\varsigma(2^{-m}-n)
\end{equation}
The time series can be represented by a forecast series on the $\xi(t)$ and $\varsigma(t)$  wavelets with multilevel analysis indexed by $k \in {0,1,2,...}$ and $j \in {0,1,2,...J}$ and can be represented as follows it is formulated:
\begin{multline}\label{s}
  x(t)=\Sigma_mS_{m,n}\xi_{m,n}(t)+\Sigma_Kd_{m,n}\varsigma_{m,n}(t)+\\
  \Sigma_md_{m-1,n}\varsigma_{m-1,n}(t)+...+\Sigma_md_{m,n}\varsigma_{m,n}(t)
\end{multline}

Where the expansion coefficients $s_{(m,n)}$ and $d_{(m,n)}$  are given by the following:

\begin{equation}\label{d}
  s_{m,n}=\int \xi_{m,n}TS(t)dt
\end{equation}

\begin{equation}\label{d}
  d_{m,n}=\int \varsigma_{m,n}TS(t)dt
\end{equation}

The multi-scale approximation of the time series $TS(t)$ is presented as follows:
\begin{equation}\label{s}
  SL_m(t)=\Sigma_ns_{m,n}\xi_{m,n}(t)
\end{equation}
\begin{equation}\label{d}
  JL_m(t)=\Sigma_Kd_{m,n}\varsigma_{m,n}(t)
\end{equation}
Orthogonal wavelet series approximation can be shown as follows:
\begin{equation}\label{d}
  x(t)=SL_m(t)+JL_m(t)+JL_{n-1}(t)+...+JL_1(t)
\end{equation}

\subsection{Stacked model}
The general purpose of this model is to provide a learning representation that can use several models in presenting the final results. Each model predicts the input data and the final output is obtained from the average prediction of the models. For example, if $P_1$ is the prediction result of model 1, $P_2$ is the prediction result of model 2, and $P_n$ is the prediction result of model $n$, the overall output will be as follows:
\begin{equation}\label{d}
  P_{end}=\frac{P_1+P_2+P_3+...+P_n}{n}
\end{equation}
While the majority vote was used for the classification model.

Stacked models have been reviewed in the result section because comprehensive research has been done on traditional machine learning models, the review of these models has been omitted here, and only deep learning approaches have been used. Some of the most important models used in this study have been discussed in the following. These models are used in a stacked form, as shown in Figure \ref{CNS}.

We have three types of RNN: Long short-term memory (LSTM) cells \cite{R1ee}, gated recurrent units (GRUs) \cite{R2F} and independent RNN(IndRNN). Various versions of these units exist in the literature, so we briefly summarize the ones used here.

\begin{itemize}
  \item \emph{\textbf{LSTM}}: A standard LSTM cell contains three gates: the forgetting gate($f_t$), which determines how much of the previous data will be forgotten; An input gate($i_t$) that evaluates the information to be written to the cell memory. Furthermore, the output gate($o_t$) that determines how to calculate the output from the current information\cite{R1ee}:
\begin{equation}\label{df}
  f_t=\sigma(Wei_fx_t+Uei_fh_{t-1}+bei_f)
\end{equation}
\begin{equation}\label{df}
  i_t=\sigma(Wei_ix_t+Uei_ih_{t-1}+bei_i)
\end{equation}
\begin{equation}\label{df}
  o_t=\sigma(Wei_ox_t+Uei_oh_{t-1}+bei_o)
\end{equation}
\begin{equation}\label{df}
  c_t=f_t\circ c_{t-1}+i_t\circ(Wei_cx_t+Uei_ch_{t-1}+bei_c)
\end{equation}
\begin{equation}\label{f}
  h_t=o_t\circ \tau(c_t)
\end{equation}

where $n$ size of input, $m$ size of cell state and output, $x_t$ input vector(time t, size $n*1$), $f_t$ forget gate vector($m*1$), $i_t$ input gate vector($m*1$), $O_t$ output gate vector(size=$m*1$), $h_t$ output vector (size=$m*1$), $c_t$ cell state vector(size=$m*1$), $[Wei_f, Wei_i, Wei_o, Wei_c]$ input gate weight matrices (size=$m*n$), $[Uei_f, Uei_i, Uei_o, Uei_c]$ output gate weight matrices (size=$m*m$), $[bei_f, bei_i, bei_o, bei_c]$ bias vectors(size=$m*1$), $\sigma$ logistic sigmoid activation function, and $\tau$ tanh activation function.

\begin{equation}\label{s}
  L=\lambda_1L_{seg}+\lambda_2L_{DC}
\end{equation}

\item \textbf{GRU network:} GRUs  use fewer parameters and  update ($u_t$) and reset ($rt$) gates. Gate $u_t$ sets the update rate of the hidden state while gate $r_t$ decides how much of the past information to forget by resetting parts of the memory\cite{RR33}. The following set of equations defines the GRU unit\cite{RR33}.
      \begin{equation}\label{d}
        z_t=\sigma_g(Wei_zx_t+U_zh_{t-1}+b_z)
      \end{equation}
      \begin{equation}\label{d}
        r_t=\sigma_g(Wei_rx_t+U_rh_{t-1}+b_r)
      \end{equation}
       \begin{equation}\label{d}
       \hat{h}_t=\phi_h(Wei_hx_t+U_h(r_t\odot h_{t-1})+b_h)
      \end{equation}
      \begin{equation}\label{d}
       h_t=z_t\odot \hat{h}_t+(1-z_t)\odot h_{t-1}
      \end{equation}

      where $x_t$ input vector, $h_t$ output vector, $\hat{h}_t$ candidate activation vector, $z_t$ update gate vector, $r_t$ rest gate vector ,  $[Wei, U, b] $ parameter matrices and vector, $\sigma_g$ sigmoid function, and $\phi_h$ hyperbolic tangent.
   \item \textbf{IndRNN:}   IndRNN was proposed in \cite{RE3} as a main component of RNN. It's defined as follows:
       \begin{equation}\label{d}
         hiden-vector_t=\sigma(Wx_t+u\bigodot hiden vector_{t-1}+b)
       \end{equation}
       $x_t\in R^M$ and $hiden-vector_t\in R^N$ are the input stated and hidden stated in time step $t$, respectively. $w\in R^{N*M}$, $u\in R^N$, and $b\in R^N$ respectively show the current input, return input, and bias weights; $\odot$ represents the Hadamard product, $\sigma$ is an essential activation function like Relu, sigmoid, tanh, or etc. which is Relu in this paper. Each neuron of IndRNN in a layer is independent of others, and the correlation between neurons is checked by superimposing two or more layers.
Since the neurons are independent, the gradient propagation over time can be calculated for each neuron separately. Since the neurons are independent, the gradient propagation over time can be calculated for each neuron separately. For the $n-th$ neuron, the gradient is calculated as follows:
\begin{equation}\label{s}
\frac{\delta J_{n,T}}{\delta h_{n,t}}=\frac{\delta J_{n},T}{\delta h_{n,T}}u_n^{T-1}\Pi_{k=T}^{T-1}\sigma'_{n,k+1}
\end{equation}
Please refer to \cite{RE3} for more details on the above derivation.
\end{itemize}

\subsection{Transformer}

In this section, we will first introduce the transformer network. The linear transformer (LT) \cite{L1T} is a variant of the well-known transformer architecture widely used for NLP and other sequence modelling tasks. The primary difference between the linear transformer and the baseline transformer is in the way the mechanism of attention is applied to a sequence.

The transformer model was proposed by the Google team and published in 2017 for NLP and other sequential problems\cite{L3T}. This model was proposed to overcome time series problems in traditional RNNs. This model was proposed to learn relationships at different locations in the input sequence and to better capture long dependencies in the time series.
\begin{figure*}[!hptb]
\begin{center}
\includegraphics[width=1\textwidth]{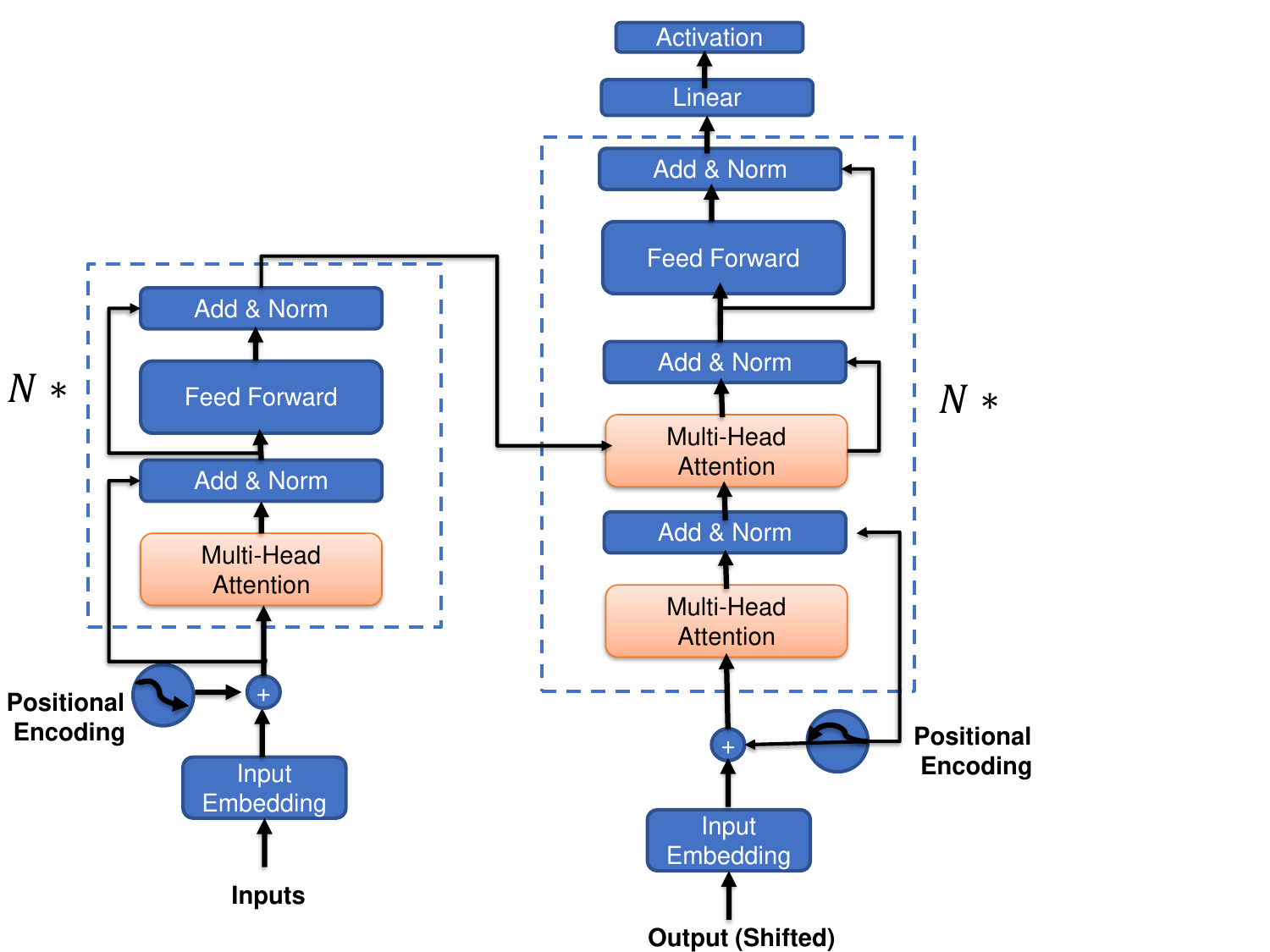}
\caption{The transformer model.}
\label{fikdsd1s}
\end{center}
\end{figure*}
As shown in Figure \ref{fikdsd1s}, the proposed transformer-based framework consists of two main modules: (1) The first module is the encoder module, which can receive long sequence input. (2) The final second is the decoder module, which receives long sequence inputs, zeros out the target elements, measures the weighted attention composition of the feature map, and instantly predicts the output elements generatively.
\subsubsection{Input embedding }
RNN-based models use a time pattern with a repeating structure, while the transformer uses a point-attention mechanism, and timestamps act as local location context. To obtain long-range dependencies, global information such as hierarchical timestamps is required. First, we project the scalar field $x_{ti}$ to the vector $d$ model-dim with an LSTM network. Second, we use fixed-position embedding to represent the local texture.

\begin{equation}
PE_{(pos,2j)}=sin(\frac{pos}{2L_x})^{\frac{2j}{d_{model}}}
\end{equation}

\begin{equation}
PE_{(pos, 2j+1)}=cos(\frac{pos}{2L_x})^{\frac{2j}{d_{model}}}
\end{equation}

$d_{model}$ represents the feature dimension after input embedding, and $Lx$ is the length of the input.
The self-attention score was found to have a high "scatter", meaning that some dot-product pairs could contribute to the attention mechanism, and others could be ignored \cite{zhou2021informer}. The i-m query attention can be defined as a kernel in the following possible form:

\begin{equation}
Attention(q_i, K, V)=\sum\frac{k(q_i,k_j)}{\sum k(q_i,k_l)}v_j
\end{equation}

In this regard, the attention of query $i$ to all keys is defined as probability $p(k_j|q_i$, and the output is its combination with $v$ values.  We want to identify the most important queries that can be achieved by measuring the similarity between p and q using the Kullback-Leibler divergence. The dispersion measure of query $i$ can be defined as follows:
\begin{equation}
    M(q_i,K)=ln\sum_{j=1}^{l_k}e^{\frac{q_iK^T}{
\sqrt{2}}}-\frac{1}{L_K}\sum_{j=1}^{l_k}e^{\frac{q_iK^T}{
\sqrt{2}}}
\end{equation}
In the following, the  ProbSparse Self-attention can be defined as:
\begin{equation}
    Attention(Q,K,V)=softmax(\frac{\hat{Q}K^T}{\sqrt{d}})V
\end{equation}

Given an input sequence of tokens represented as a matrix $X$ with dimensions $(\text{sequence\_length}, \text{embedding\_dimension})$, the attention mechanism calculates a set of attention scores $A$ as follows:

For each position $i$ in the sequence:

\begin{enumerate}
    \item Generate three new matrices:
        \begin{itemize}
            \item Query matrix $Q_i$ by multiplying $X_i$ (the input at position $i$) with a learnable weight matrix $W_Q$.
            \item Key matrix $K_i$ by multiplying $X$ with another learnable weight matrix $W_K$.
            \item Value matrix $V_i$ by multiplying $X$ with a third learnable weight matrix $W_V$.
        \end{itemize}

    \item Compute the attention scores between the query $Q_i$ and all key positions in the sequence using the dot product:
    \[
    A_i = \text{softmax}\left(\frac{Q_i K^T}{\sqrt{d_k}}\right)
    \]
    \begin{itemize}
        \item $A_i$ represents the attention scores for position $i$.
        \item $Q_i$ is the query matrix for position $i$.
        \item $K^T$ is the transpose of the key matrix.
        \item $d_k$ is the key vectors' dimension, typically a fraction of the embedding dimension.
    \end{itemize}

    \item Use the attention scores to compute a weighted sum of the value matrices:

    \[
    O_i = A_i V
    \]

    \begin{itemize}
        \item $O_i$ represents the output (context) vector for position $i$.
        \item $A_i$ is the attention scores for position $i$.
        \item $V$ is the set of value matrices for all positions.
    \end{itemize}

\end{enumerate}

The attention scores reflect how much each position's information contributes to the representation of the current position. This mechanism allows the model to focus more on relevant parts of the input when making predictions or encoding information.

In summary, the attention mechanism calculates attention scores by comparing queries with keys to measure the importance of different positions in the input sequence. This process captures relationships and dependencies, enabling the model to understand context and relationships within the data.

The attention mechanism can be mathematically represented as:

\[
\text{Attention}(Q, K, V) = \text{softmax}\left(\frac{QK^T}{\sqrt{d_k}}\right)V
\]

The Attention Calculation Formula is as follows:

Given:

\begin{align*}
    Q & : \text{Query matrix} \\
    K & : \text{Key matrix} \\
    V & : \text{Value matrix} \\
    d_k & : \text{Dimension of keys} \\
    n & : \text{Number of elements in the sequence}
\end{align*}

The attention score $(score)$ between a query element $(q_i)$ and a key element $(k_j)$ is calculated as:

\[
score(q_i, k_j) = q_i \cdot k_j^T
\]

To improve the weighting, the attention scores are often scaled by the square root of the dimension of keys $(d_k)$:

\[
scaled\_score(q_i, k_j) = \frac{score(q_i, k_j)}{\sqrt{d_k}}
\]

The scaled scores are then passed through a softmax function to get the attention weights $(w_{ij})$:

\[
w_{ij} = \frac{e^{scaled\_score(q_i, k_j)}}{\sum_{j=1}^{n} e^{scaled\_score(q_i, k_j)}}
\]

These attention weights represent the importance of each key element $(k_j)$ with respect to the query element $(q_i)$. They are then used to compute a weighted sum of the value elements $(v_j)$ to obtain the final output:

\[
attention(q_i, K, V) = \sum_{j=1}^{n} w_{ij} \cdot v_j
\]

\section{MATERIAL}

Here, we briefly describe the preparatory material we used for developing the Deep Stack architecture. In particular, we focus on the BTC dataset used for evaluating our system and the basic tools used for building our model.

\begin{figure*}[!hptb]
\centering
\begin{center}
\includegraphics[width=1\textwidth]{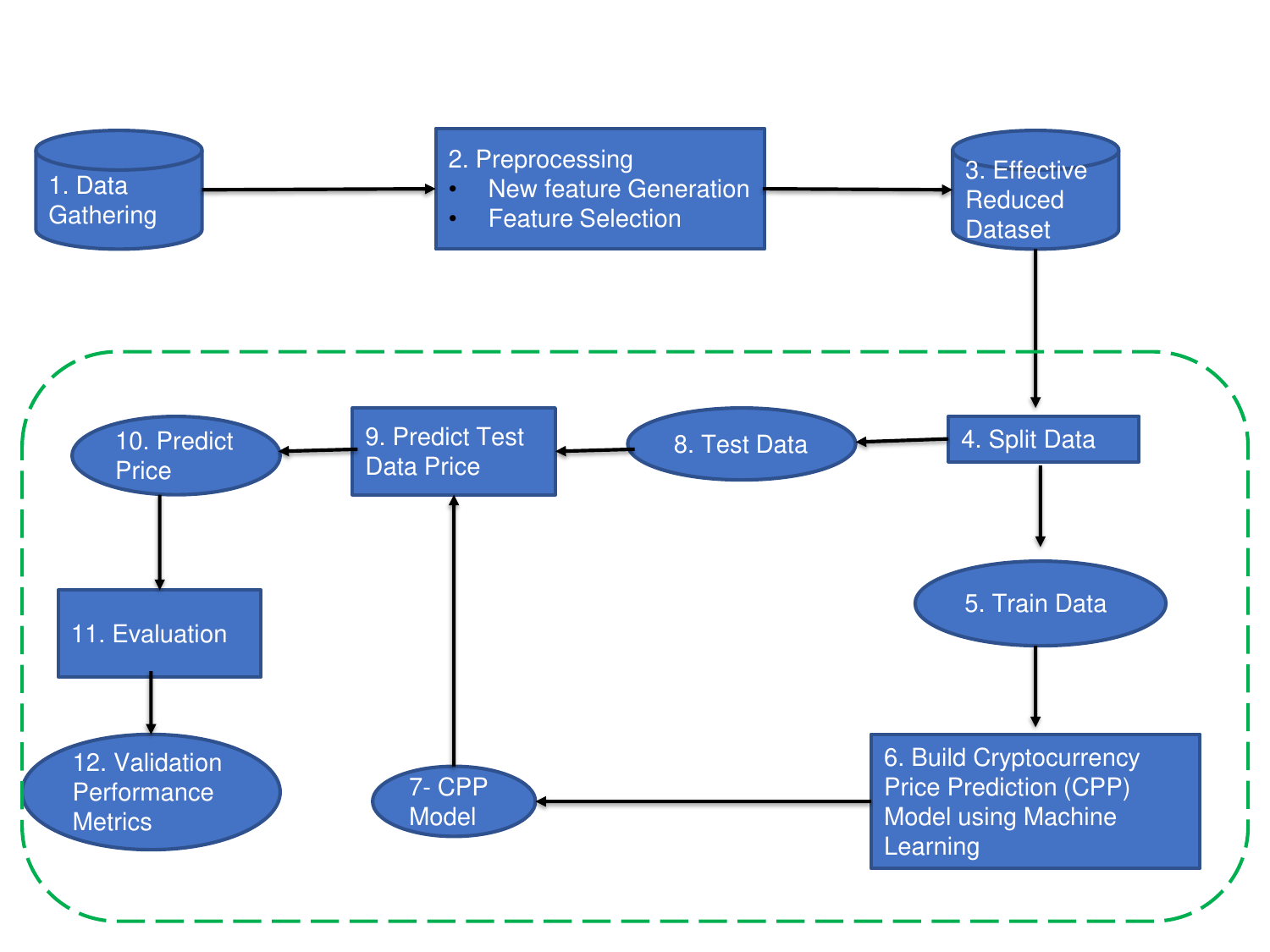}
\caption{Price prediction and classification model. }
\label{CPPK}
\end{center}
\end{figure*}

\subsection{The Bitcoin Dataset}
To predict the price of Bitcoin, in Figure\ref{CPPK}, based on the literature review, a model is presented that includes the following steps:
\begin{itemize}
 \item \textbf{Dataset collection:} The data set used in this research includes BTC hash rate information, which is discussed in detail in the data collection section.
\item \textbf{Pre-processing:} This step is usually the pre-processing of the hash rate data, which includes the extraction of indicator features and feature selection algorithms.
\item \textbf{Reduced or golden dataset}: This data set can greatly improve the accuracy of price prediction and is generally used for training and evaluating models.
\item \textbf{Data division:} The data-sharing policy of the evaluation training is located in this section. Depending on the policy, validation data can also be considered in this section.
\item \textbf{Train data: } Specifies the training dataset(80\% of the total data).
\item \textbf{ Building a ML  model:} At this stage, a ML algorithm is needed that interacts with training data and can predict future prices with high accuracy.
\item \textbf{CPP model:}After training the built learning model, it can be used to predict the price of the next stage, which is classified and regression.
\item \textbf{Test data:} This part of data is used to evaluate the built model(20\% of the total data).
\item \textbf{Predict the Test Data Price: }The built model predicts the price of the test data partition.
\item \textbf{Predicted price:} The output of the CPP model on test data.
\item \textbf{Evaluation:} The predicted CPP is compared with the test data's actual values, and the performance criteria provided in this step are calculated.
\item \textbf{Credit performance criteria:} The price tag is a numerical value, so the difference between the predicted and actual value will be more significant. The calculation criteria are as follows:

\begin{equation}\label{s}
MAE=\frac{1}{n}\sum_{i=1}^n|\eta_i-\varsigma_i|
\end{equation}

\begin{equation}\label{s}
RMSE=\sqrt{\frac{1}{n}\sum_{i=1}^n|\eta_i-\varsigma|^2}
\end{equation}

\begin{equation}\label{s}
MAPE=\frac{100}{n}\sum_{i=1}^n\frac{|\eta_i-\varsigma_i|}{\eta_i}
\end{equation}

where $\eta$ represents actual labels and $\varsigma$ represents predicted labels. low MAE, MAPE, and RMSE is desirable.

In the classification problem, the labels are 0 or 1. For this purpose, the following evaluation criteria are used. Accuracy, F-1 score, and Area Under Curve (AUC) metrics are used to evaluate the performance of classification models. A model with
high Accuracy, AUC, and F-1 score is desirable.

\begin{equation}\label{s}
  Actual_{Accuracy}=\frac{(\sharp_{TP}+\sharp_{TN})}{(\sharp_{TP}+\sharp_{TN}+\sharp_{FP}+\sharp_{FN})}
\end{equation}

\begin{equation}\label{s}
Actual_{Precision}=\frac{\sharp_{TP}}{\sharp_{TP}+\sharp_{FP}}
\end{equation}

\begin{equation}\label{s}
Actual_{Recall}=\frac{\sharp_{TP}}{\sharp_{TP}+\sharp_{FN}}
\end{equation}

\begin{equation}\label{s}
F1-score=2*\frac{Actual_{precision}*Actual_{recall}}{Actual_{precision}+Actual_{recall}}
\end{equation}

\begin{equation}\label{s}
Specificity=\frac{\sharp_{TN}}{\sharp_{TN}+\sharp_{FP}}
\end{equation}

\end{itemize}
At first, the BTC data set, which includes BTC hash rate information, was collected. BTC features and price data are available online for free. The data of this study was collected from bitinfocharts\footnote{\url{https://bitinfocharts.com}}. 712 features were collected based on technical indicators. We used the same data collection process as in \cite{mudassir2020time}. The feature selection method selected a smaller subset of features from this large set of features. Technical indicators:
\begin{itemize}
\item Simple Moving Average:$SMA=\frac{(A_1+A_2+...+A_n)}{A}$. Where $(A_1+A_2+...+A_n)$ are prices, and $n$ is The number of total periods.
\item Exponential Moving Average: $EMA=Price(t)*K*EMA(y)*(1-k)$. Where $t$ is today, $y$ is yesterday, $k=\frac{2}{N+1}$, and $N$ number of days in EMA.
\item  Relative Strength Index: $RSI=100-[\frac{100}{\frac{n_{up}}{n_{down}}}]$. $n_{up}$ is the average of n-day up closes and $n_{down}$ is the average of n-day down closes.  This article uses 15 days RSI.
\item Weighted Moving Average: $WMA=\frac{Price_1*n+Price_2*(n-1)+...+Price_n}{\frac{n*(n+1)}{2}}$. Where $n$ is the time period.
\item STD: Standard Deviation
\item VAR: Variance
\item Triple Exponential Moving Average:$TRIX=\frac{MA_i^3-MA_{i-1}^3}{MA_{i-1}^3}*100$. Where $MA^3=$MA(period) of $MA^2$, $MA^2=$MA(period) of $MA$, and $MA=$MA(period) of $close$
\item Rate Of Change: $ROC=(\frac{Closing price_p-Closing price_{p-n}}{Closing price_{p-n}})$. where $Closing price_{p}$ is closing price of most recent period and $Closing price_{p-n}$ is Closing price $n$ periods befor most recent period.
\end{itemize}
were used for these data. The extracted hash rate information includes the following:

\begin{itemize}
\item Tr: Number of BTC payments sent and received.
\item MTF: Median transaction fee in BTC.
\item ATF: Each transaction in the sending process can have an associated fee that the sender determines. Transactions with higher fees are given higher priority, encouraging Bitcoin miners to process them sooner than transactions with lower fees.
\item ATV: Average value of transactions price in BTC
\item MTV: Median value of transactions price in BTC.
\item BS: Transaction information cryptographically linked in the blockchain. The maximum block size is currently set to 1 MB.
\item SFA: These are the distinct BTC addresses from which payments are made daily.
\item BT: The time required to mine a block.
\item Di: Average daily mining difficulty. T
\item HA: The total daily computing capacity of the BTC network. The hash rate shows the speed of the computer in completing an operation.
\item MP: profitability in dollars/day for one THash per second (THash/s).
\item Active addresses: the number of unique addresses participating in a transaction by sending or receiving BTC.
\item BitS: Total BTC sent daily.
\item T100: The ratio of BTC stored in the top 100 accounts to all other BTC accounts.
\item FTRR: The ratio of the fee sent in a transaction to the reward for confirming that transaction by other users.
\end{itemize}

\subsubsection{Preprocessing}
In preprocessing, the linear interpolation method imputes missing value items as much as possible.  20\%  of the data was kept for validation, and 80\%  was used for training. On the other hand, the Isolation Forest method algorithm \cite{IO} was used to control outliers.

\subsubsection{Feature Selection}
Feature selection(FS)  have become an essential part of the learning in machine and deep models for dealing with high-dimensional features. Selecting suitable features can improve the inductive learner in learning speed, generalization capacity, and model simplicity\cite{IO3}.
Dimensionality reduction methods are often divided into FS and feature extraction(FE)\cite{IO4}. On the one hand, feature extraction methods reduce dimensionality by combining key features. Hence, they can construct a set of new features that are usually more compact and distinct. These methods are preferred in applications such as image analysis, image processing and information retrieval because model accuracy is more important than its interpretability \cite{IO4}.  The principles of FS will be explained below.

FS is an important part of preprocessing in data-driven systems that improves the performance of learning models. In the FS process, features are repeatedly extracted and pruned by different approaches after extraction. First, feature importance is determined using an ensemble method based on a random decision forest. In the second step, it examines the reduced feature set for multicollinearity and cross-correlation. The subset of the resulting features has relatively high importance with low cross-correlation values and no multicollinearity. FS is repeated for each of the three data collection intervals. This extraction method is called the Embedded feature. This approach was used with two other FS approaches: Chi2 and RFE. In the following, we will examine two other approaches to feature selection:
\begin{equation}\label{s}
  ni_j=w_j C_j-w_{left(j)} C_{left(j)} -w_{right(j)}  C_{right(j)}
\end{equation}
where $ni_j$ is the importance of node $j$ and $w_j$ is the weighted number of samples that reach node $j$. $C_j$ is the impurity value of node $j$. $left(j)$ is the child node of the left division on node $j$. $right(j)$ is the child node of the right division on node $j$. Then, the importance of each feature in the decision tree is obtained based on the following relationship \cite{CHY}:

\begin{equation}\label{s}
  fi_i=\frac{\Sigma_{j:node j splits on feature i}ni_j}{\Sigma_{k\in all nodes}ni_j}
\end{equation}
Where $fi_i$ is the importance of feature $i$. $ni_j$ is the importance of node $j$.
Then these values can be normalized by dividing them by the sum of feature importance values\cite{CHY}:
\begin{equation}\label{s}
  norm fi_i=\frac{fi_i}{\Sigma_{j\in all features}fi_i}
\end{equation}
Finally, the feature importance of a feature at the random forest(RF) is calculated by the average of all trees in the following relationship\cite{CHY}:
\begin{equation}\label{s}
  RF fi_i=\frac{\Sigma_{j:all tree} norm fi_{ij} }{T}
\end{equation}
$RF fi_i$ is the importance of feature $i$ in the RF model. $norm fi_{ij}$ is the normalized feature importance score of feature $i$ in trained tree $j$. $T$ is the total number of trees in the RF model.

Chi-squared Statistics method evaluates the relationship between two categorical variables. For this purpose, numerical variables must be divided into several intervals. The Chi-square statistic is obtained as follows\cite{SCHI}:
\begin{equation}\label{s}
  X^2=\Sigma_{i=1}^m\Sigma_{j=1}^k(\frac{A_{ij}-E_{ij}}{E_{ij}})
\end{equation}
Where $m$, $k$, and $A_{ij}$ are the number of intervals, classes, and samples of the $i-th$ interval of class $j$, respectively. $E_{ij}=R_i*\frac{C_j}{N}$  where the number of samples in the i-th interval, $C_j$ is the number of samples in class $j$, and $N$ is the total number of samples and the expected frequency of $A_{ij}$. Next, the features selected for each interval are listed in Table \ref{T1YYd}.
\begin{table*}[]
\tiny
\centering
\caption{The selected features by different feature selection methods for each interval, where $*$ values indicate the selection of that feature in the intervals.}
\begin{tabular}{|l|ccc|ccc|ccc|}
\hline
 & \multicolumn{3}{c|}{Embedded} & \multicolumn{3}{c|}{FRE} & \multicolumn{3}{c|}{Chi2} \\ \hline
intervals & \multicolumn{1}{c|}{90} & \multicolumn{1}{c|}{30} & 7 & \multicolumn{1}{c|}{90} & \multicolumn{1}{c|}{30} & 7 & \multicolumn{1}{c|}{90} & \multicolumn{1}{c|}{30} & 7 \\ \hline
MTF30(median transaction for fee trx(30) USD) & \multicolumn{1}{c|}{*} & \multicolumn{1}{c|}{*} &  & \multicolumn{1}{c|}{*} & \multicolumn{1}{c|}{*} &  & \multicolumn{1}{c|}{*} & \multicolumn{1}{c|}{*} &  \\ \hline
MTF7(median transaction for fee trx(7) USD) & \multicolumn{1}{c|}{*} & \multicolumn{1}{c|}{} &  & \multicolumn{1}{c|}{*} & \multicolumn{1}{c|}{} & * & \multicolumn{1}{c|}{} & \multicolumn{1}{c|}{} & * \\ \hline
P90EMA (price  ema(90) per USD) & \multicolumn{1}{c|}{*} & \multicolumn{1}{c|}{} &  & \multicolumn{1}{c|}{*} & \multicolumn{1}{c|}{} &  & \multicolumn{1}{c|}{*} & \multicolumn{1}{c|}{} &  \\ \hline
S90(size  trx(90)) & \multicolumn{1}{c|}{*} & \multicolumn{1}{c|}{*} & * & \multicolumn{1}{c|}{*} & \multicolumn{1}{c|}{*} & * & \multicolumn{1}{c|}{*} & \multicolumn{1}{c|}{} & * \\ \hline
Tran(transactions) & \multicolumn{1}{c|}{*} & \multicolumn{1}{c|}{} &  & \multicolumn{1}{c|}{*} & \multicolumn{1}{c|}{} & * & \multicolumn{1}{c|}{*} & \multicolumn{1}{c|}{} & * \\ \hline
P30w(price wma(30) USD) & \multicolumn{1}{c|}{} & \multicolumn{1}{c|}{*} &  & \multicolumn{1}{c|}{*} & \multicolumn{1}{c|}{} & * & \multicolumn{1}{c|}{*} & \multicolumn{1}{c|}{} &  \\ \hline
P3w(price wma(3) USD) & \multicolumn{1}{c|}{} & \multicolumn{1}{c|}{*} & * & \multicolumn{1}{c|}{*} & \multicolumn{1}{c|}{*} &  & \multicolumn{1}{c|}{*} & \multicolumn{1}{c|}{*} &  \\ \hline
P7(price wma(7) USD) & \multicolumn{1}{c|}{} & \multicolumn{1}{c|}{} & * & \multicolumn{1}{c|}{*} & \multicolumn{1}{c|}{*} & * & \multicolumn{1}{c|}{*} & \multicolumn{1}{c|}{*} & * \\ \hline
MTF7R(median transaction for fee roc(7) USD) & \multicolumn{1}{c|}{} & \multicolumn{1}{c|}{} & * & \multicolumn{1}{c|}{*} & \multicolumn{1}{c|}{} & * & \multicolumn{1}{c|}{*} & \multicolumn{1}{c|}{} & * \\ \hline
D30R(difficulty rsi(30)) & \multicolumn{1}{c|}{*} & \multicolumn{1}{c|}{*} & * & \multicolumn{1}{c|}{} & \multicolumn{1}{c|}{*} & * & \multicolumn{1}{c|}{} & \multicolumn{1}{c|}{} & * \\ \hline
MP(mining profitability) & \multicolumn{1}{c|}{*} & \multicolumn{1}{c|}{*} & * & \multicolumn{1}{c|}{*} & \multicolumn{1}{c|}{*} & * & \multicolumn{1}{c|}{*} & \multicolumn{1}{c|}{*} & * \\ \hline
P30S(price sma(30) USD) & \multicolumn{1}{c|}{} & \multicolumn{1}{c|}{*} & * & \multicolumn{1}{c|}{*} & \multicolumn{1}{c|}{*} & * & \multicolumn{1}{c|}{*} & \multicolumn{1}{c|}{*} & * \\ \hline
S90E(sent in usd ema(90) USD) & \multicolumn{1}{c|}{} & \multicolumn{1}{c|}{*} & * & \multicolumn{1}{c|}{} & \multicolumn{1}{c|}{*} & * & \multicolumn{1}{c|}{} & \multicolumn{1}{c|}{*} &  \\ \hline
TVU (transaction value USD) & \multicolumn{1}{c|}{*} & \multicolumn{1}{c|}{*} & * & \multicolumn{1}{c|}{*} & \multicolumn{1}{c|}{*} & * & \multicolumn{1}{c|}{*} & \multicolumn{1}{c|}{*} & * \\ \hline
T100C(top 100 cap) & \multicolumn{1}{c|}{*} & \multicolumn{1}{c|}{} & * & \multicolumn{1}{c|}{*} & \multicolumn{1}{c|}{*} & * & \multicolumn{1}{c|}{*} & \multicolumn{1}{c|}{*} & * \\ \hline
D90M(difficulty mom(90)) & \multicolumn{1}{c|}{*} & \multicolumn{1}{c|}{*} &  & \multicolumn{1}{c|}{} & \multicolumn{1}{c|}{*} &  & \multicolumn{1}{c|}{} & \multicolumn{1}{c|}{*} &  \\ \hline
H90V(hashrate var(90)) & \multicolumn{1}{c|}{*} & \multicolumn{1}{c|}{*} &  & \multicolumn{1}{c|}{*} & \multicolumn{1}{c|}{*} &  & \multicolumn{1}{c|}{*} & \multicolumn{1}{c|}{*} &  \\ \hline
P90W(price wma(90) USD) & \multicolumn{1}{c|}{*} & \multicolumn{1}{c|}{*} &  & \multicolumn{1}{c|}{*} & \multicolumn{1}{c|}{*} & * & \multicolumn{1}{c|}{*} & \multicolumn{1}{c|}{*} & * \\ \hline
S90S(sent in usd sma(90) USD) & \multicolumn{1}{c|}{*} & \multicolumn{1}{c|}{*} &  & \multicolumn{1}{c|}{*} & \multicolumn{1}{c|}{*} &  & \multicolumn{1}{c|}{*} & \multicolumn{1}{c|}{*} &  \\ \hline
MTF (median transaction fee USD) & \multicolumn{1}{c|}{*} & \multicolumn{1}{c|}{*} &  & \multicolumn{1}{c|}{*} & \multicolumn{1}{c|}{*} & * & \multicolumn{1}{c|}{*} & \multicolumn{1}{c|}{} & * \\ \hline
\end{tabular}
\label{T1YYd}
\end{table*}

This data set consists of 3 different intervals. Table \ref{cahpt52} shows the range of each of these intervals.

\begin{table*}[!htbp]
\tiny
\renewcommand{\arraystretch}{1.3}
\caption{Number of training and test samples in different intervals.}
\label{cahpt52}
\centering
\tiny
\begin{tabular}{|c|c|c|c|}
\hline
Dataset	&Interval I&	Interval II&	Interval III\\		
\hline
Range&	[ 2013/04/01 - 2016/04/01]&	[2013/04/01 - 2017/04/01]&	[2013/04/01 - 2019/12/31]	\\
\hline
Number of Records	&1,206	&1,462&	2,466	\\
\hline
Train data size (80\%)&	964	&1,169&	1,972	\\
\hline
Test data size (20\%)	&242&	293&	494\\
\hline
\end{tabular}
\end{table*}

\section{EVALUATION}
The deep stack approach discussed in Section III has been evaluated by adopting the BTC data set. Here, we describe the implemented validation procedure and we discuss the obtained results.
\subsection{Evaluation Procedure}

The model validation procedure used 80\% to 20\% adjustment for training and testing data. For this purpose, the proposed model was compared with several basic models, which are listed below:

\begin{itemize}
  \item \textbf{ANN}\cite{mudassir2020time}: The neural network considered by the authors includes hyperparameters:
   \begin{itemize}
     \item Optimizer=Adam
     \item 2(128,128) Hidden layers(neurons)
     \item learning rate(LR)=0.08
     \item 500 Epoch
     \item Batch size(BS)=64
     \item Activation Function(AF)=Relu
     \item Loss function(LF)=logcosh
   \end{itemize}
 The basic article used This network in two regression and classification modes.
  \item \textbf{SANN)}\cite{mudassir2020time}: In this approach, 5 ANN networks were considered with the settings mentioned in the ANN approach. SANN consists of 5 individual ANNs that are used to train a larger ANN model. Individual models are trained using training data with 5-fold cross-validation; each model is trained with the same configuration in a separate layer. Since ANNs have random initial weights, each trained ANN gets different weights, and this advantage enables them to learn their differences well. \cite{mudassir2020time} used this network in two modes
      : regression and classification.
  \item \textbf{SVM}\cite{mudassir2020time}: This algorithm is a supervised ML model that operates based on the idea of separating points using a hyperplane, and in fact, its primary goal is to maximize the margin. In SVM, kernels can be linear or non-linear depending on the data, including radial basis function (RBF), tangent hyperbola, and polynomial. This algorithm can provide predictions with a low error rate for small data sets without much training. The \cite{mudassir2020time} considers this approach with Gaussian RBF kernel for classification and regression problems.
  \item \textbf{LSTM}\cite{mudassir2020time}: This approach is an RNN network that uses four gates to learn long sequences. In the previous section, RNN approaches were discussed. This approach is used in both regression and classification modes, and in this research, its regression mode was used depending on the type of labels.
  \item \textbf{1D-CNN+IndRNN}\cite{Na}: In this model, the authors tried combining 1D-CNN and IndRNN neural networks to extract spatial and temporal features. For this purpose, they used a 1D-CNN layer at the beginning of the network. This network provided spatial features to an IndRNN layer. At the end of their proposed network, two neurons were used for classification, and one neuron was used for classification.
\end{itemize}

\subsection{Price Classification}
At first, classification issues were discussed. For this purpose, the price of the next step was set to 1 in case of increase and 0 in case of decrease. Due to having two classes, network architectures were considered binary classifications. Table \ref{Table1} shows the results obtained from different feature selection approaches along with the deep stack model. $\pm$ values indicate the offset of the results in different execution rounds. In interval I, the Chi2 feature selection approach reached an accuracy of $62\pm1$, the highest accuracy among the tested models. This approach also reached $F1-score=0.73\pm0.02$ and $AUC=0.54\pm 0.02$. RFE and Embedded approaches worked the same in this interval. These two approaches were able to achieve $accuracy=1\pm61$, $F1=0.74\pm0.01$ and $AUC=0.55\pm0.01$. In interval II, the Embedded approach recorded higher results. This approach was able to achieve $accuracy=1\pm62$, $F1=0.72\pm0.02$, and $AUC=0.55\pm0.01$. Chi2 and RFE approaches also achieved an accuracy of $60\pm1$ and $61\pm1$, respectively. In interval III, the best result was obtained by the Chi2 approach. This approach was able to achieve $accuracy=57\pm1$, $F1=0.53\pm0.01$ and $AUC=0.50\pm0.01$. The two RFE and Embedded approaches achieved $53\pm1$ accuracy in these data. The forecasts were slightly different in the 7, 30, and 90-day forecasts. In 7 days, the RFE approach achieved $accuracy=66\pm1$, $F1=0.62\pm0.01$ and $AUC=0.65\pm0.01$, the best result among the tested approaches. The two approaches, Chi2 and Embedded in this interval, reached the accuracy $63\pm1$. In 30 days, the Embedded approach achieved the best result. This approach achieved $accuracy=68\pm1$, and two Chi2 and RFE approaches achieved an accuracy of $66\pm1$ and $67\pm1$, respectively. In 90 days, the proposed approach recorded the best result on Chi2. This approach was able to achieve $accuracy=81\pm1$, $F1=0.77\pm0.02$, and $AUC=0.72\pm0.02$.
%\usepackage{multirow}
\begin{table}[]
\tiny
\label{Table1}
\caption{The results were obtained by the deep stack approach in combination with feature selection approaches (classification problem).}
\begin{tabular}{|l|l|l|l|l|}
\hline
Metrics & Intervals & Chi2 & RFE & Embedded \\ \hline
\multirow{-1}{*}{Acc.\%} & I & 62$\pm1$ & 61$\pm1$ & 61$\pm1$ \\ {2-5}
 & II & 60$\pm1$ & 61$\pm1$ & 62$\pm1$ \\ {2-5}
 & III & 57$\pm1$ & 53$\pm1$ & 53$\pm1$ \\ {2-5}
 & 7 & 63$\pm1$ & 66$\pm1$ & 63$\pm1$ \\ {2-5}
 & 30 & 66$\pm1$ & 67$\pm1$ & 68$\pm1$ \\ {2-5}
 & 90 & 81$\pm1$ & 71$\pm1$ & 63$\pm1$ \\ {2-5}
 & Avg \% & 64.83 & 63.16 & 61.66 \\ \hline
\multirow{-1}{*}{F1-score} & I & 0.73$\pm0.02$ & 0.74$\pm0.01$ & 0.74$\pm0.01$ \\ {2-5}
 & II & 0.73$\pm0.02$ & 0.61$\pm0.01$ & 0.72$\pm0.02$ \\ {2-5}
 & III & 0.66$\pm0.01$ & 0.53$\pm0.01$ & 0.52$\pm0.01$ \\ {2-5}
 & 7 & 0.51$\pm0.02$ & 0.62$\pm0.01$ & 0.54$\pm0.01$ \\ {2-5}
 & 30 & 0.66$\pm0.02$ & 0.60$\pm0.01$ & 0.66$\pm0.01$ \\ {2-5}
 & 90 & 0.77$\pm0.02$ & 0.71$\pm0.01$ & 0.60$\pm0.01$ \\ {2-5}
 & Avg \% & 0.67 & 0.63 & 0.63 \\ \hline
\multirow{-1}{*}{AUC} & I & 0.52$\pm0.01$ & 0.52$\pm0.01$ & 0.52$\pm0.01$ \\ {2-5}
 & II & 0.55$\pm0.01$ & 0.55$\pm0.01$ & 0.55$\pm0.01$ \\ {2-5}
 & III & 0.52$\pm0.01$ & 0.50$\pm0.01$ & 0.52$\pm0.01$ \\ {2-5}
 & 7 & 0.59$\pm0.01$ & 0.65$\pm0.01$ & 0.59$\pm0.01$ \\ {2-5}
 & 30 & 0.71$\pm0.01$ & 0.67$\pm0.01$ & 0.71$\pm0.01$ \\ {2-5}
 & 90 & 0.72$\pm0.01$ & 0.73$\pm0.01$ & 0.72$\pm0.01$ \\ {2-5}
 & Avg \% & 0.61 & 0.60 & 0.60 \\ \hline
\end{tabular}
\end{table}

The highest average accuracy(AVG\%) was obtained by the Chi2 feature selection. This approach was able to reach AVG=64.83. Also, in average F1 and AUC, this approach had the highest performance and could reach Average F1=0.67 and Average AUC=0.62.

\subsection{Price Prediction}

In the regression problems, it was tried to consider the price of BTC as a predictor. For this purpose, only one neuron was considered in the last layer of the pre-institutional network. Table \ref{Table2} shows the results obtained from the proposed approach using feature selection approaches. In interval I, the Chi2 feature selection approach reached $MAE=1.21\pm0.01$, $RMSE=1.85\pm0.02$, and $MAPE=0.50\pm0.02$, while the RFE approach reached $MAE=1.31\pm0.02$, $RMSE=1.89\pm0.02$, and $MAPE=0.58\pm0.02$, and the Embedded approach also reached $MAE=1.28\pm0.03$, $RMSE = 1.87\pm0.02$, AND $MAPE = 0.58\pm0.02$. In interval II, the Chi2 approach reached $MAE=6.48\pm0.02$, $RMSE=10.18\pm0.2$, and $MAPE=1.47\pm0.02$. The RFE approach, which had the best result in this interval, achieved $MAE=5.25\pm0.02$, $RMSE=8.17\pm0.5$, and $MAPE=1.44\pm0.02$. The Embedded approach also reached $MAE=1.28\pm0.03$, $RMSE=1.87\pm0.02$, and $MAPE=0.58\pm0.02$ in this interval.
In interval III, the Chi2 approach could reach $MAE=49.52\pm0.02$, $RMSE=79.20\pm1.2$, and $MAPE=2.10\pm0.02$. The RFE approach was able to reach $MAE=38.35\pm0.02$, $RMSE=79.20\pm1.2$ and $MAPE=1.94\pm0.02$, and the Embedded approach achieved better results in this interval and was able to achieve $MAE=37.81\pm0.02$, $RMSE=75.59\pm2.00$ and $MAPE=2.08\pm0.02$.
On other forecasting days, different results were obtained. In 7-day forecasting, the best MAE, RMSE, and MAPE were obtained by the Chi2 approach. In the 30-day interval, the best MAE was obtained by the RFE approach, the best RMSE by the Embedded approach, and the best MAPE by the Chi2 approach. In the 90-day interval, the Embedded approach achieved lower MAE and RMSE and obtained the best result. In this interval, the best MAPE was obtained by the Chi2 approach.

\begin{table}[]
\tiny
\label{Table2}
\caption{The results were obtained by the deep stack approach in combination with feature selection approaches (regression problem).}
\begin{tabular}{|l|l|l|l|l|}
\hline
Metrics & Intervals & Chi2 & RFE & Embedded \\ \hline
\multirow{-1}{*}{MAE} & I & 1.21$\pm0.01$  & 1.31$\pm0.02$  & 1.28$\pm0.03$  \\ {2-5}
 & II & 6.48$\pm0.02$  & 5.25$\pm0.02$ & 6.01$\pm0.01$ \\ {2-5}
 & III & 49.52$\pm0.02$ & 38.35$\pm0.02$ & 37.81$\pm0.02$ \\ {2-5}
 & 7 & 13.86$\pm1.01$ & 16.87$\pm0.1$ & 17.36$\pm0.2$ \\ {2-5}
 & 30 & 89.57$\pm0.02$ & 88.96$\pm2.00$ & 103.29$\pm2.00$ \\ {2-5}
 & 90 & 70.81$\pm1.1$ & 97.03$\pm1.00$ & 82.33$\pm1.00$ \\ {2-5}
 & Avg\% & 38.57 & 41.29 & 41.34 \\ \hline
\multirow{-1}{*}{RMSE} & I & 1.85$\pm0.02$ & 1.89$\pm0.02$ & 1.87$\pm0.02$ \\ {2-5}
 & II & 10.18$\pm0.02$ & 8.17$\pm0.05$ & 8.95$\pm0.02$ \\ {2-5}
 & III & 93.27$\pm0.05$ & 79.20$\pm1.2$ & 75.59$\pm2.00$ \\ {2-5}
 & 7 & 25.35$\pm0.05$ & 35.97$\pm2.4$ & 37.93$\pm1.01$ \\ {2-5}
 & 30 & 189$\pm0.02$ & 213$\pm3.00$ & 178$\pm0.02$ \\ {2-5}
 & 90 & 121$\pm0.02$ & 192.24$\pm3$ & 162.10$\pm0.02$ \\ {2-5}
 & Avg\% & 73.44 & 88.41 & 77.40 \\ \hline
\multirow{-1}{*}{MAPE} & I & 0.58$\pm0.02$ & 0.55$\pm0.00$ & 0.58$\pm0.02$ \\ {2-5}
 & II & 1.55$\pm0.02$ & 1.44$\pm0.00$ & 1.55$\pm0.02$ \\ {2-5}
 & III & 2.08$\pm0.02$ & 1.94$\pm0.02$ & 2.08$\pm0.02$ \\ {2-5}
 & 7 & 2.93$\pm0.02$ & 2.85$\pm0.02$ & 2.93$\pm0.02$ \\ {2-5}
 & 30 & 3.79$\pm0.02$ & 3.73$\pm0.02$ & 3.79$\pm0.02$ \\ {2-5}
 & 90 & 2.72$\pm0.02$ & 3.89$\pm0.02$ & 3.72$\pm0.02$ \\ {2-5}
 & Avg\% & 2.27 & 2.4 & 2.34 \\ \hline
\end{tabular}
\end{table}

In the average mode, the RFE approach obtained the lowest MAE value of 41.29. In the average RMSE mode, the Embedded approach obtained the lowest value, equal to 77.40. Moreover, finally, the Chi2 approach obtained the lowest MAPE value, equal to 2.23.
\subsection{Comparison with base models }
Table 9 shows the results of the classification models for the six investigated intervals. In the interval I, deep stack performed best in the Acc evaluation criterion and reached accuracy=$62\pm1$. In this interval, the 1D-CNN+IndRNN approach also obtained better results than other comparative approaches and achieved accuracy=61. In Interval II, SANN and Transformer achieved 65\% accuracy, the highest in this interval. In interval III, the 1DCNN+IndRNN approach achieved accuracy=62, which was the highest accuracy. For 7, 30 and 90 day forecasting, the proposed deep stack approach reached the highest accuracy. This approach achieved an accuracy of $63\pm1$, $66\pm1$ and $81\pm1$, respectively. The proposed approach achieved Average accuracy=64.83, Average F1-score=0.67, and Average AUC=0.61 in average mode. The 1D-CNN+IndRNN approach performed better than our proposed approach in the average F1 and AUC mode and achieved Average F1-score=0.71 and Average AUC=0.62. In general, the highest classification accuracy was 81\%, obtained by the proposed approach and in the 90-day forecast. Also, the highest f1-score obtained by the predictive approach was obtained in the 90-day prediction and was equal to 0.79. However, the highest AUC was obtained by the 1D-CNN+IndRNN approach, which was equal to 0.74.

\begin{table*}[]
\centering
\tiny
\caption{Performance of classification models and proposed Deep Stack in different intervals.}
\begin{tabular}{|l|l|c|c|c|c|c|c|c|c|}
\hline
Metrics & Intervals & ANN & SVM & SANN & LSTM & 1DCNN+indRNN & GRU & Transformer & Deep Stack\\ \hline
\multirow{-1}{*}{Acc.\%} & I & 57 & 55 & 60 & 54 & 61 & 56 & 60 & 62$\pm1$  \\ {2-10}
 & II & 56 & 62 & 65 & 53 & 64 & 62 & 65 & 60$\pm1$ \\ {2-10}
 & III & 53 & 56 & 60 & 54 & 62 & 57 & 61 & 57$\pm1$ \\ {2-10}
 & 7 & 51 & 62 & 53 & 55 & 60 & 61 & 62 & 63$\pm1$ \\ {2-10}
 & 30 & 52 & 52 & 62 & 52 & 63 & 60 & 61 & 66$\pm1$ \\ {2-10}
 & 90 & 62 & 54 & 60 & 64 & 73 & 62 & 68 & 81$\pm1$ \\ {2-10}
 & Avg\% & 51.16 & 56.83 & 60 & 55.33 & 63.83 & 59.66 & 62.83 & 64.83 \\ \hline
\multirow{-1}{*}{F1-score} & I & 0.72 & 0.67 & 0.71 & 0.63 & 0.76 & 0.67 & 0.72 & 0.73$\pm0.02$ \\ {2-10}
 & II & 0.69 & 0.74 & 0.75 & 0.58 & 0.78 & 0.68 & 0.59 & 0.73$\pm0.01$ \\ {2-10}
 & III & 0.61 & 0.53 & 0.60 & 0.66 & 0.63 & 0.62 & 0.66 & 0.66$\pm0.02$ \\ {2-10}
 & 7 & 0.65 & 0.58 & 0.38 & 0.65 & 0.68 & 0.61 & 0.67 & 0.51$\pm0.02$ \\ {2-10}
 & 30 & 0.68 & 0.68 & 0.70 & 0.68 & 0.69 & 0.67 & 0.67 & 0.66$\pm0.02$ \\ {2-10}
 & 90 & 0.56 & 0.66 & 0.68 & 0.70 & 0.73 & 0.57 & 0.69 & 0.77$\pm0.02$ \\ {2-10}
 & Avg\% & 0.65 & 0.64 & 0.63 & 0.65 & 0.71 & 0.63 & 0.66 & 0.67 \\ \hline
 \multirow{-1}{*}{AUC} & I & 0.51 & 0.51 & 0.56 & 0.52 & 0.57 & 0.57$\pm0.01$ & 0.53 & 0.52 \\ {2-10}
 & II & 0.50 & 0.55 & 0.59 & 0.53 & 0.58 & 0.56 & 0.58 & 0.55$\pm0.01$ \\ {2-10}
 & III & 0.53 & 0.56 & 0.60 & 0.54 & 0.61 & 0.58 & 0.58 & 0.52$\pm0.01$ \\ {2-10}
 & 7 & 0.53 & 0.60 & 0.51 & 0.56 & 0.61 & 0.60 & 0.60 & 0.59$\pm0.01$ \\ {2-10}
 & 30 & 0.50 & 0.50 & 0.61 & 0.50 & 0.62 & 0.57 & 0.59 & 0.71$\pm0.01$ \\ {2-10}
 & 90 & 0.59 & 0.57 & 0.62 & 0.66 & 0.74 & 0.62 & 0.69 & 0.72$\pm0.01$ \\ {2-10}
 & Avg\% & 0.52 & 0.54 & 0.58 & 0.55 & 0.62 & 0.58 & 0.59 & 0.61 \\ \hline
\end{tabular}
\end{table*}

Table \ref{FREFE} show the results of regression models for all intervals. In the interval I, BTC prices did not fluctuate much. The deep stack model reports the lowest MAE with $1.21 \pm 0.01$. The lowest MAPE was obtained by the SANN approach and reached MAPE=0.52. In the RMSE evaluation criterion, two approaches, SANN and LSTM, obtained better results.

BTC prices are significantly higher than at the end, in interval II. However, like interval I, it is relatively stable. SANN, with the lowest MAPE of 0.93, gets the best performance. The proposed approach achieved $MAPE=1.55\pm0.02$ in this interval. The transformer approach was also able to reach MAPE=1.27 in this interval.
BTC price experienced the highest volatility since April 2017, which was covered in interval III. The lowest MAPE in this interval was obtained by SVM. Nevertheless, 1D-CNN+IndRNN performed better than SVM in terms of RMSE.

In the 7-day forecast, the proposed approach obtained $MAE=13.86\pm1.01$, the lowest reported MAE. Moreover, the SVM approach obtained the worst MAE in this interval. At the 30-day interval, the proposed approach reported poorer results for 1dCNN+IndRNN. In this interval, the proposed approach reached $MAE=89.57\pm0.02$. In 90 days, the proposed approach reached $MAE=70.81\pm1.1$, $RMSE=121\pm0.02$ and $MAPE=2.72\pm0.02$, the best results among the compared and reported models.
\section{Error Analysis}

\begin{table*}[]
\centering
\tiny
\caption{Performance of Deep Stack model and other models in different intervals.}
\begin{tabular}{|l|l|l|l|l|l|l|l|l|l|}
\hline
Metrics               & Intervals & ANN    & SVM    & SANN   & LSTM   & 1DCNN+indRNN & GRU    & Transformer & Deep Stack  \\ \hline
\multirow{-1}{*}{MAE}  & I         & 4.45   & 1.72   & 1.24   & 2.20   & 1.64         & 1.71   & 1.80        & 1.21$\pm0.01$        \\ {2-10}
                      & II        & 4.61   & 5.23   & 4.13   & 6.55   & 4.35         & 5.36   & 4.30        & 6.48$\pm0.02$        \\ {2-10}
                      & III       & 39.50  & 47.04  & 40.58  & 62.90  & 37.98        & 48.01  & 40.09       & 49.52$\pm0.02$      \\ {2-10}
                      & 7         & 17.71  & 19.15  & 16.32  & 22.05  & 14.86        & 21.1   & 15.90       & 13.86$\pm1.01$     \\ {2-10}
                      & 30        & 90.12  & 87.43  & 77.12  & 116.37 & 76.8         & 91.12  & 80.01       & 89.57$\pm0.02$       \\ {2-10}
                      & 90        & 96.01  & 98.03  & 72.23  & 109.22 & 78.32        & 90.1   & 81.00       & 70.81$\pm1.1$       \\ {2-10}
                      & Avg\%     & 42.06  & 43.1   & 35.27  & 53.21  & 24           & 27.91  & 37.18       & 38.57      \\ \hline
\multirow{-1}{*}{RMSE} & I         & 6.13   & 2.37   & 1.58   & 1.58   & 2.22         & 2.45   & 2.34        & 1.85$\pm0.02$        \\ {2-10}
                      & II        & 8.22   & 9.68   & 6.48   & 10.55  & 7.97         & 9.80   & 8.00        & 10.18$\pm0.02$       \\ {2-10}
                      & III       & 74.10  & 122.34 & 87.62  & 135.76 & 79.62        & 125.89 & 79.68       & 93.27$\pm0.05$       \\ {2-10}
                      & 7         & 31.78  & 37.32  & 36.33  & 36.12  & 30.88        & 39.44  & 32.09       & 25.35$\pm0.02$       \\ {2-10}
                      & 30        & 175.60 & 158.60 & 156.30 & 219.59 & 166.99       & 160.09 & 169.90      & 189$\pm0.02$         \\ {2-10}
                      & 90        & 210.09 & 203.11 & 140.00 & 217.84 & 131.23       & 210.09 & 131.22      & 121$\pm0.02$         \\ {2-10}
                      & Avg\%     & 84.32  & 88.90  & 71.38  & 103.57 & 69.81        & 91.29  & 70.53       & 73.44      \\ \hline
\multirow{-1}{*}{MAPE} & I         & 1.88   & 0.73   & 0.52   & 0.93   & 0.68         & 1.90   & 0.72        & 0.58$\pm0.02$        \\ {2-10}
                      & II        & 1.27   & 1.23   & 0.93   & 1.98   & 1.17         & 1.43   & 1.27        & 1.55$\pm0.02$        \\ {2-10}
                      & III       & 3.78   & 1.44   & 2.73   & 3.61   & 2.06         & 1.56   & 2.29        & 2.08$\pm0.02$        \\ {2-10}
                      & 7         & 3.09   & 3.29   & 2.88   & 3.83   & 2.51         & 4.05   & 2.83        & 2.93$\pm0.02$        \\ {2-10}
                      & 30        & 4.83   & 5.50   & 3.45   & 5.96   & 3.45         & 5.60   & 3.94        & 3.79$\pm0.02$        \\ {2-10}
                      & 90        & 4.44   & 4.96   & 4.10   & 5.41   & 3.59         & 4.95   & 2.78        & 2.72$\pm0.02$        \\ {2-10}
                      & Avg\%     & 3.21   & 2.85   & 2.43   & 3.62   & 2.24         & 3.24   & 2.30        & 2.27       \\ \hline
\end{tabular}
\label{FREFE}
\end{table*}

\begin{figure*}[!htbp]
\subfloat{\includegraphics[width = 0.3\textwidth]{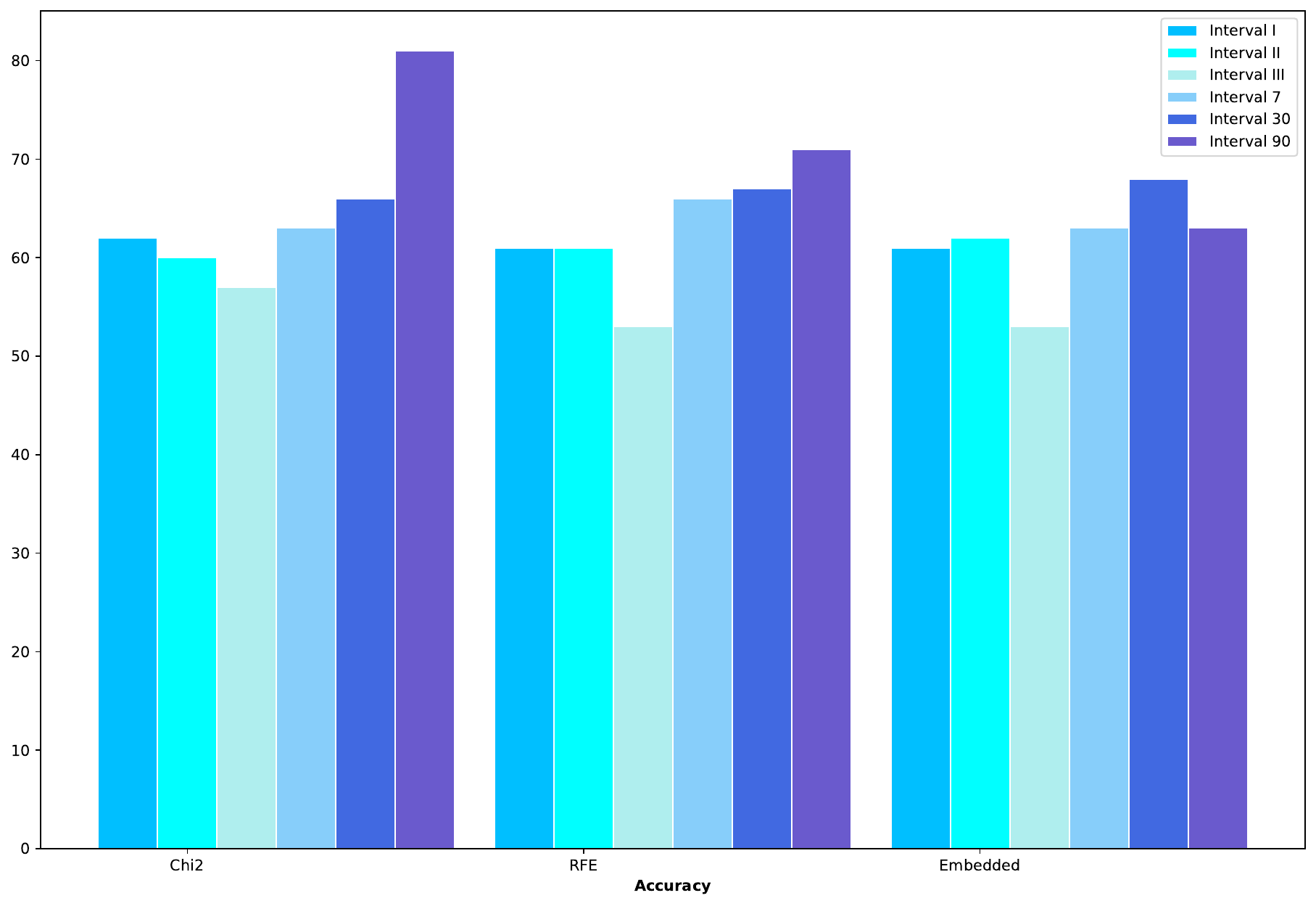}}
\subfloat{\includegraphics[width = 0.3\textwidth]{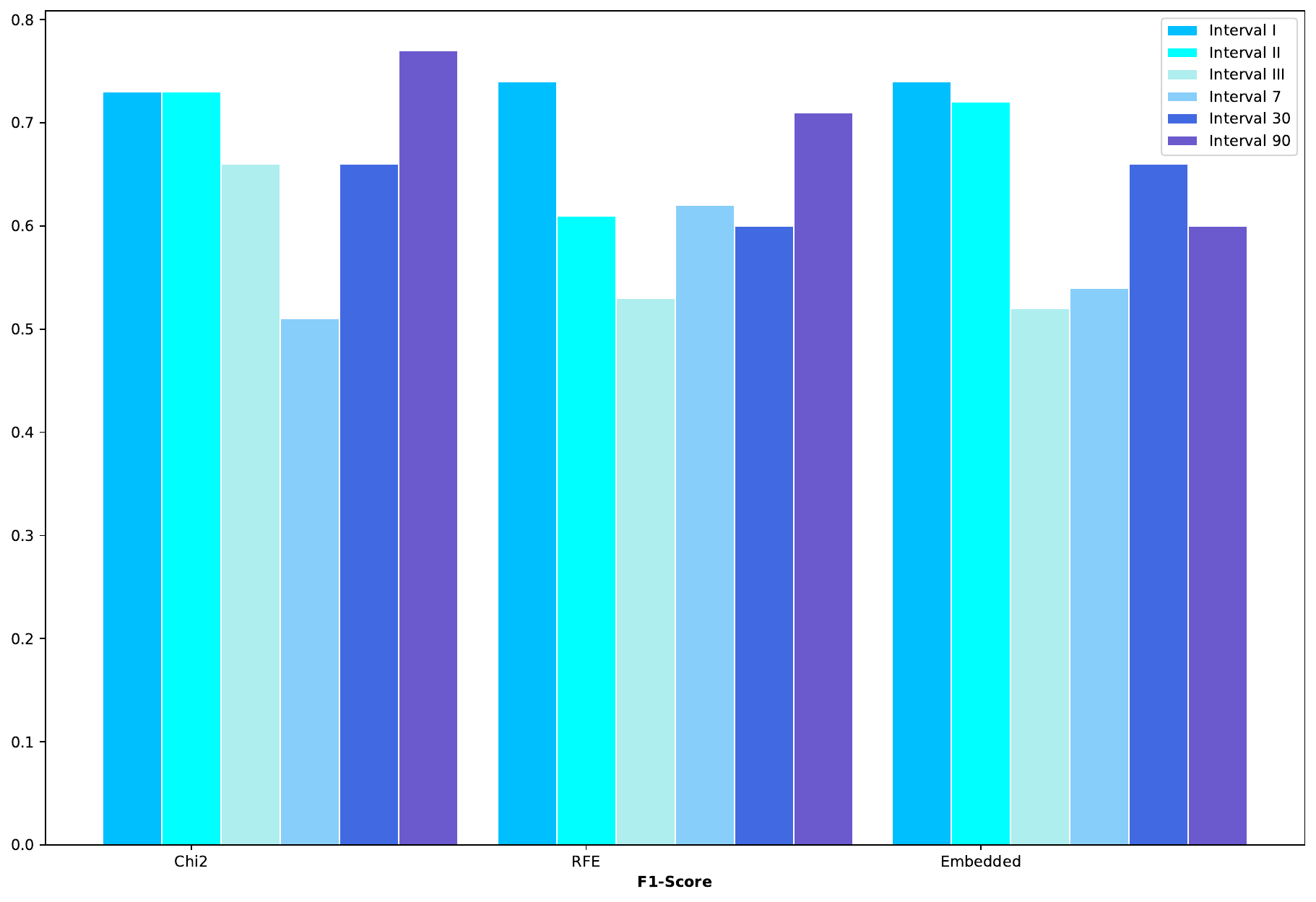}}
\subfloat{\includegraphics[width = 0.3\textwidth]{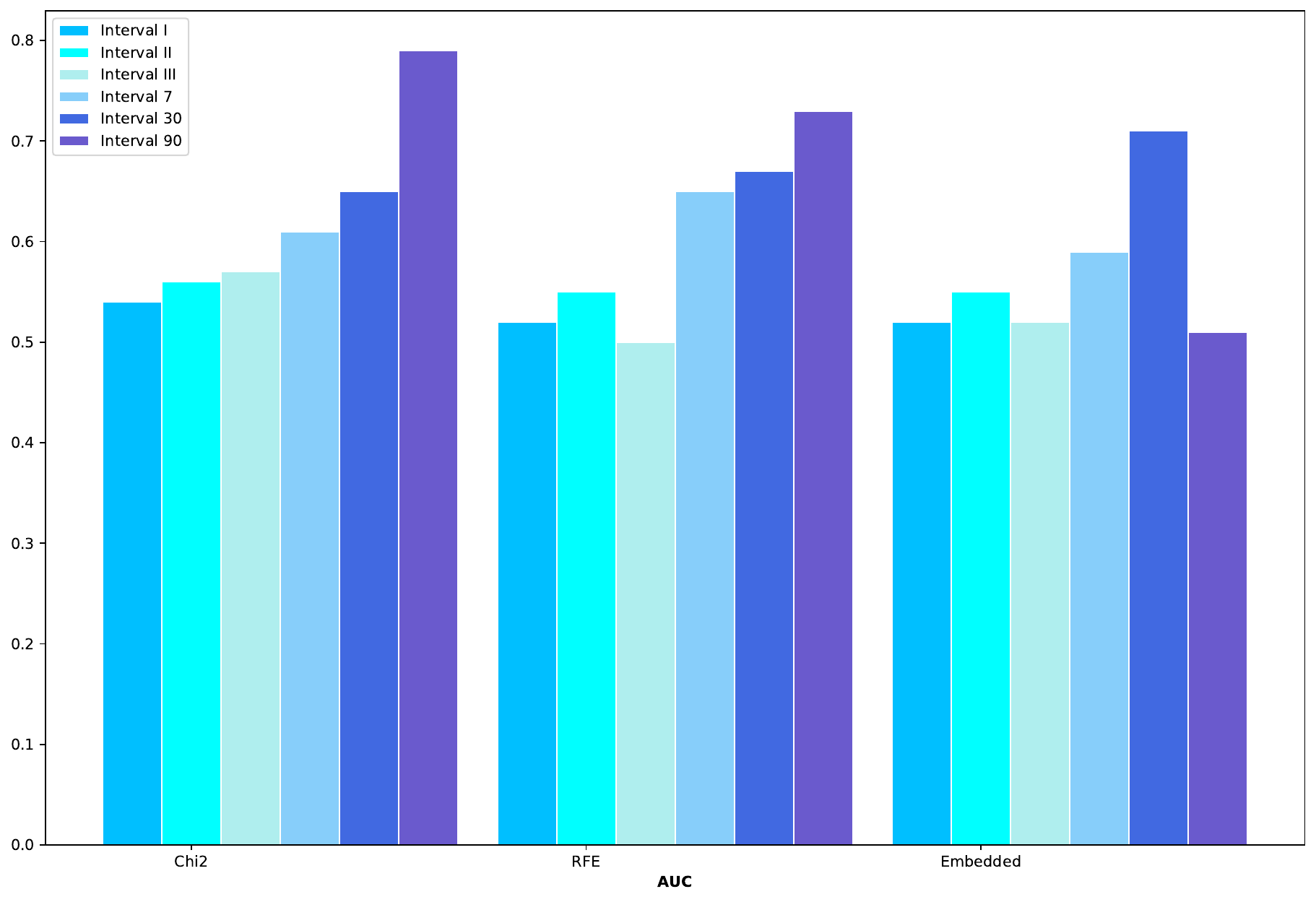}}
\caption{Accuracy, F1, and AUC of the price prediction of the Feature selection models forecast in Interval  I, II, and III.}
\label{F6}
\end{figure*}

\begin{figure*}[!htbp]
\subfloat{\includegraphics[width = 0.3\textwidth]{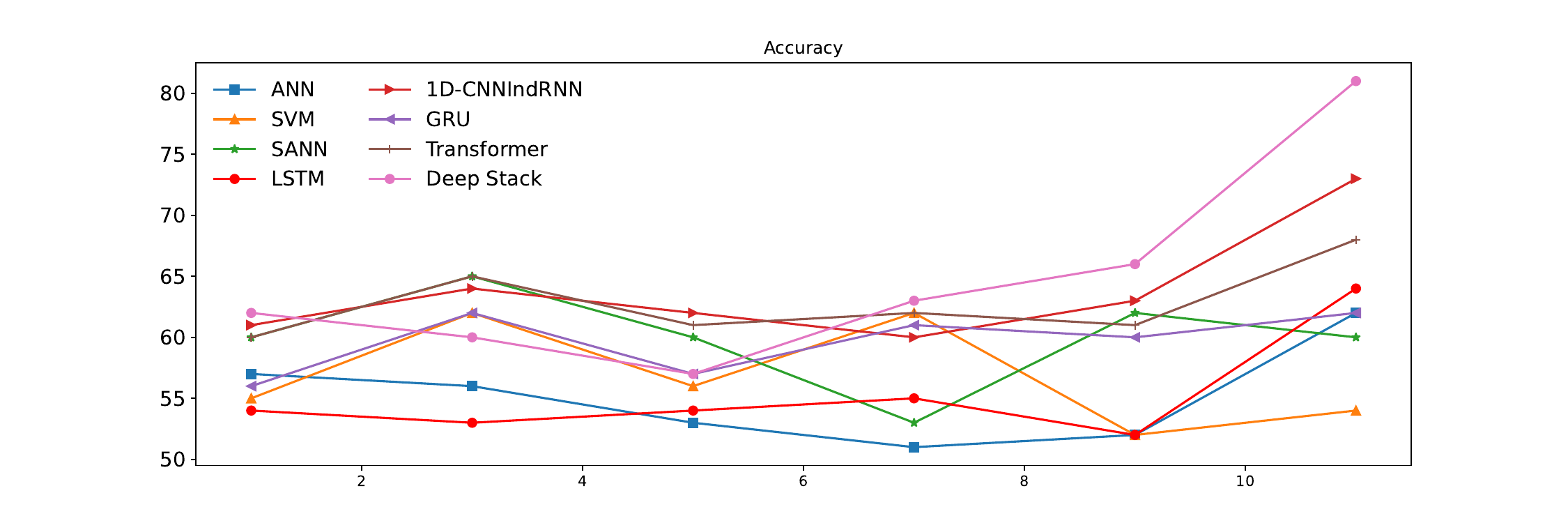}}
\subfloat{\includegraphics[width = 0.3\textwidth]{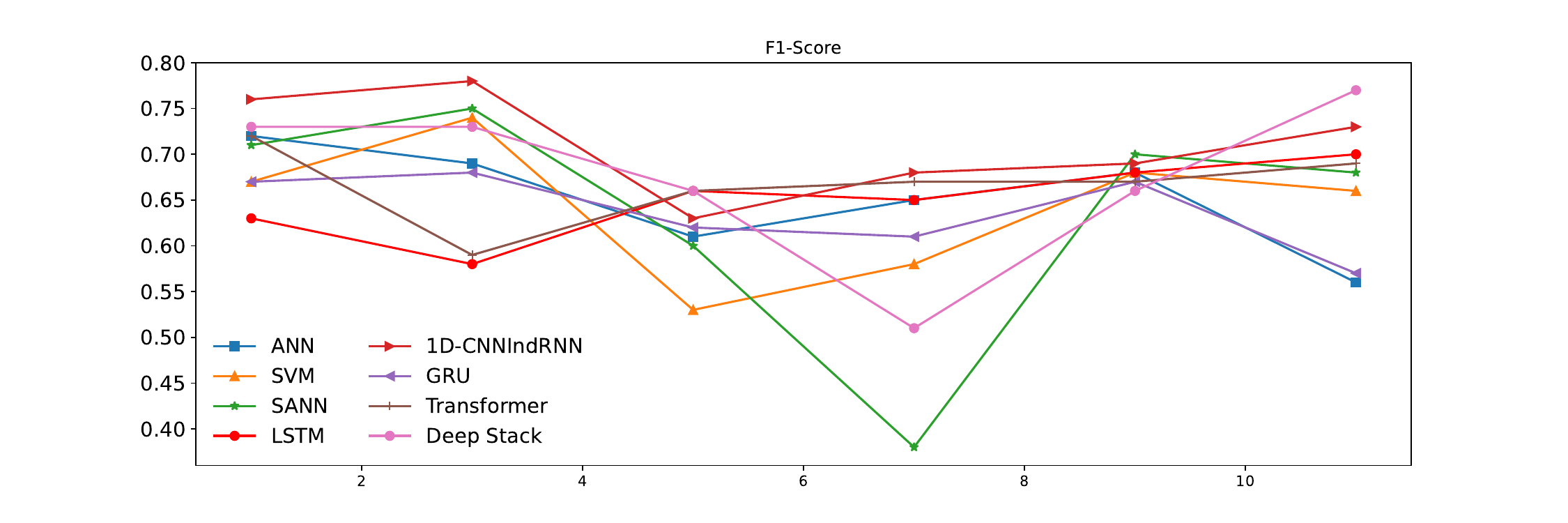}}
\subfloat{\includegraphics[width = 0.3\textwidth]{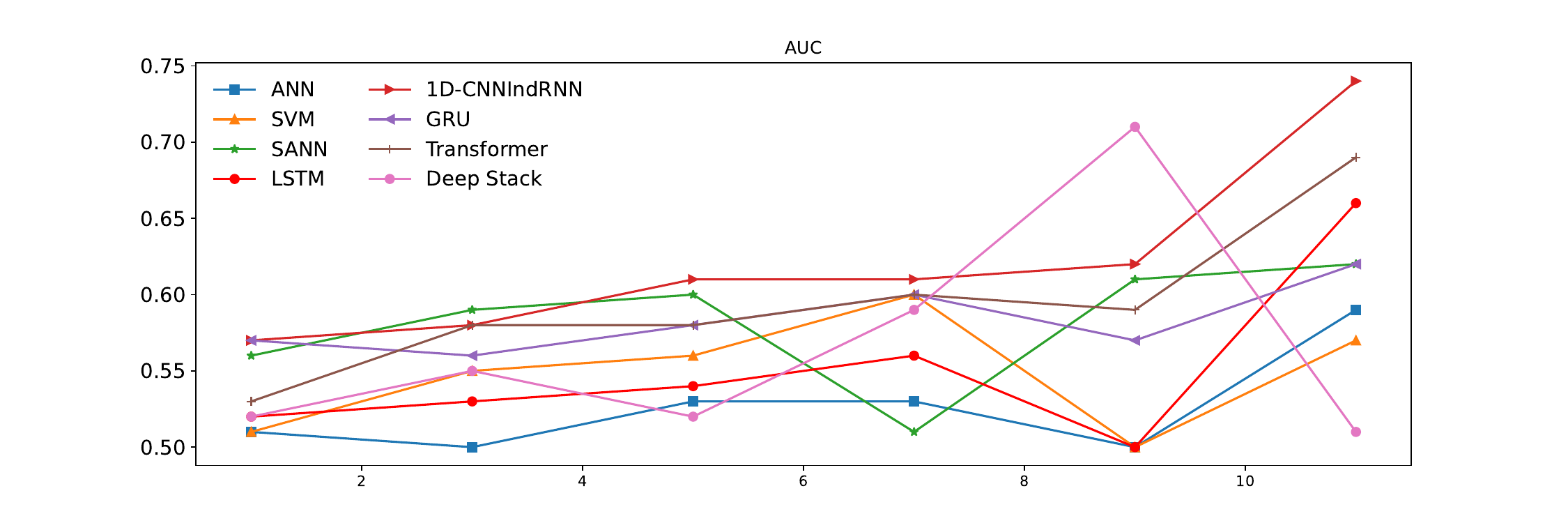}}
\caption{Accuracy, F1, and AUC of the price prediction of the proposed models and baseline for  forecast in Interval  I, II, and III.(note: $x[1]=I$,$x[3]=II$,$x[5]=III$,$x[7]=7-days$,$x[9]=30-days$,and $x[11]=90-days$) }
\label{F7}
\end{figure*}

In the error analysis, we first deal with the errors of the classification models. Figure \ref{F6} shows the Accuracy, F1-Score and AUC values for feature selection models. In these models, the Embedded and RFE approaches have obtained the worst results. The performance of these approaches could be improved in interval III. The embedded approach in 90 days has more weaknesses than other approaches. Also, Figure \ref{F7} compares the deep stack approach with other approaches in the literature.
In intervals I and II, the LSTM approach obtained the worst results; in intervals III, 7, and 30, the ANN approach obtained weaker results. Moreover, in the 90-day interval, the SVM approach obtained the weakest results.
We performed a detailed error analysis on the performance of the proposed Stacking strategy. Overall, our deep stacking model tends to provide false negative predictions. An in-depth analysis of some false predictions showed that feature representations in false negatives prevented this approach from providing better classification in these samples.

\begin{figure*}[!htbp]
\subfloat{\includegraphics[width = 0.3\textwidth]{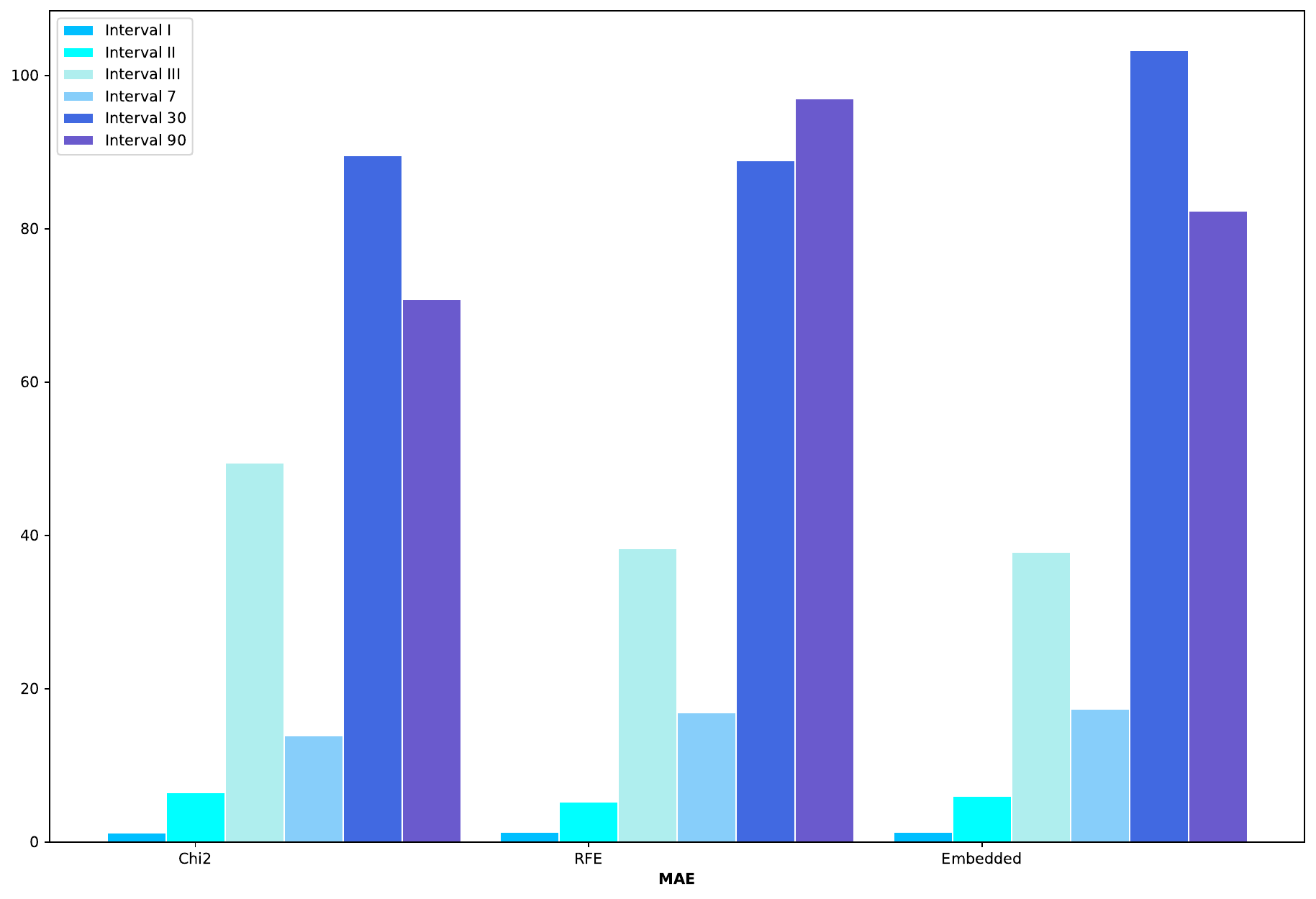}}
\subfloat{\includegraphics[width = 0.3\textwidth]{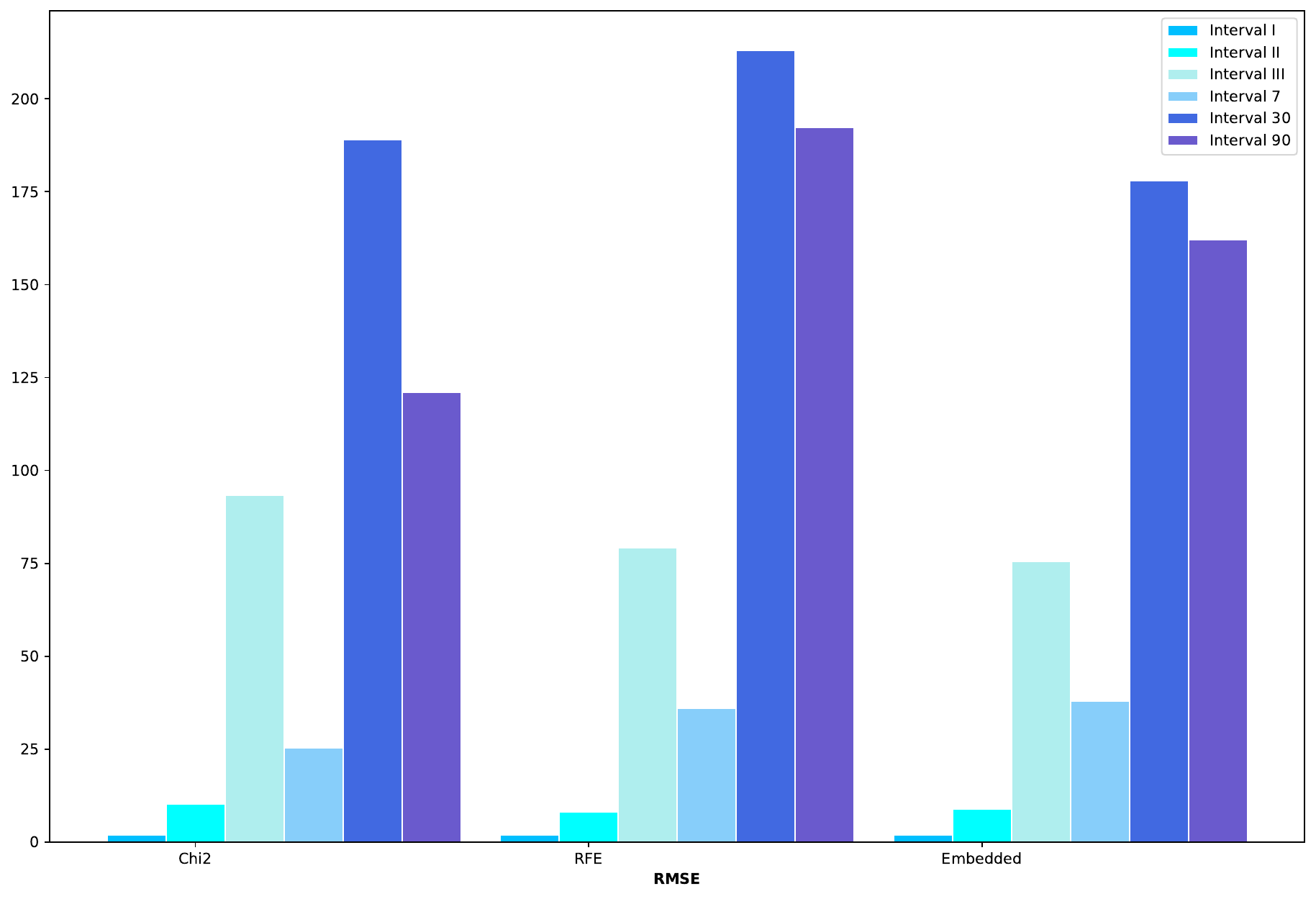}}
\subfloat{\includegraphics[width = 0.3\textwidth]{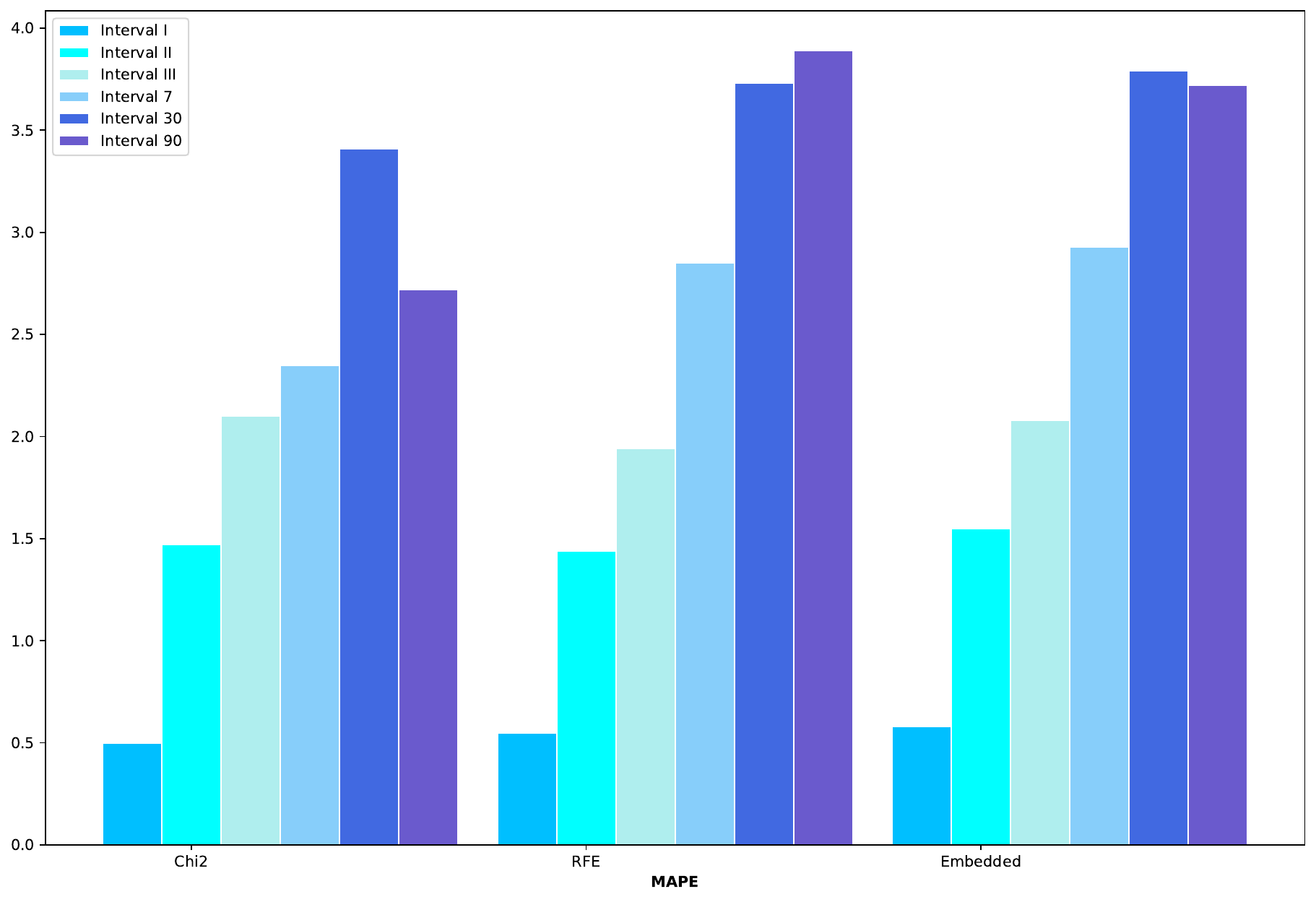}}
\caption{MAE, RMSE, and MAPE of the price prediction of the Feature selection models forecast in Interval  I, II, and III.}
\label{F6d}
\end{figure*}

\begin{figure*}[!htbp]
\subfloat{\includegraphics[width = 0.3\textwidth]{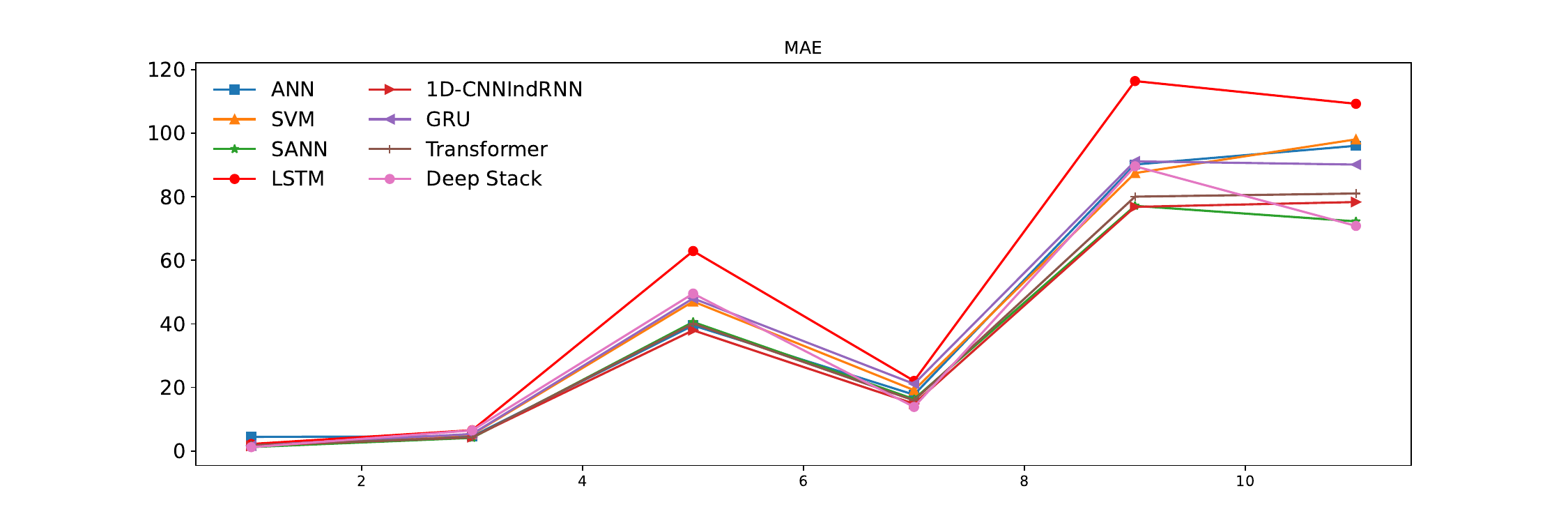}}
\subfloat{\includegraphics[width = 0.3\textwidth]{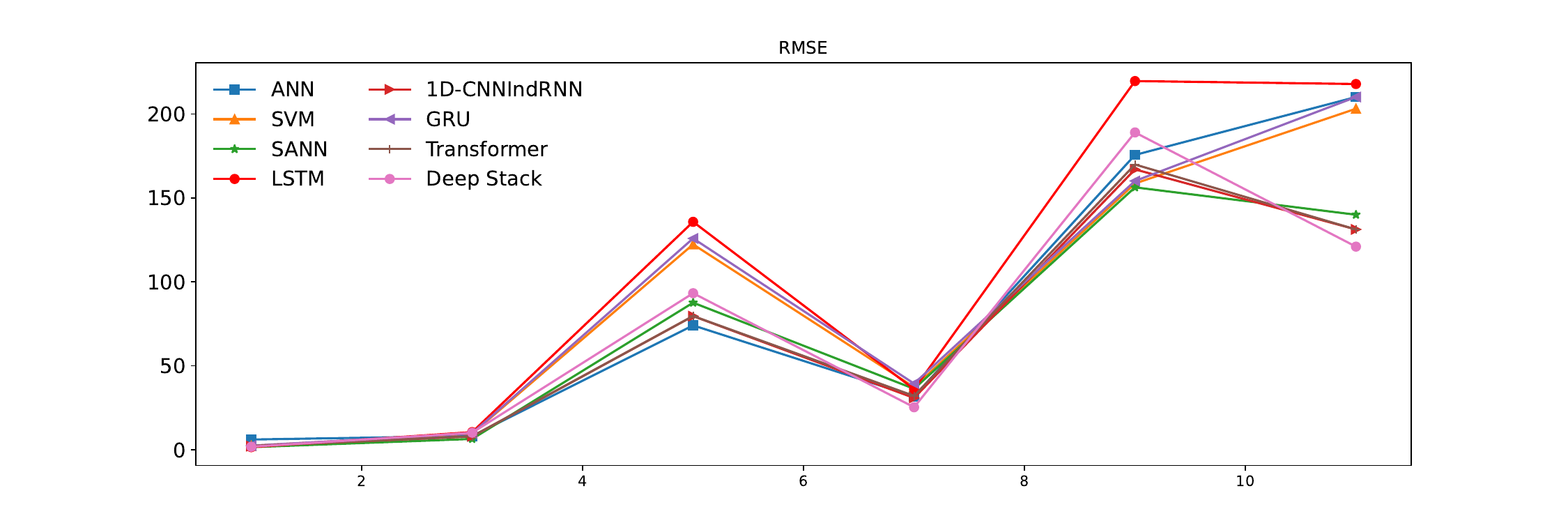}}
\subfloat{\includegraphics[width = 0.3\textwidth]{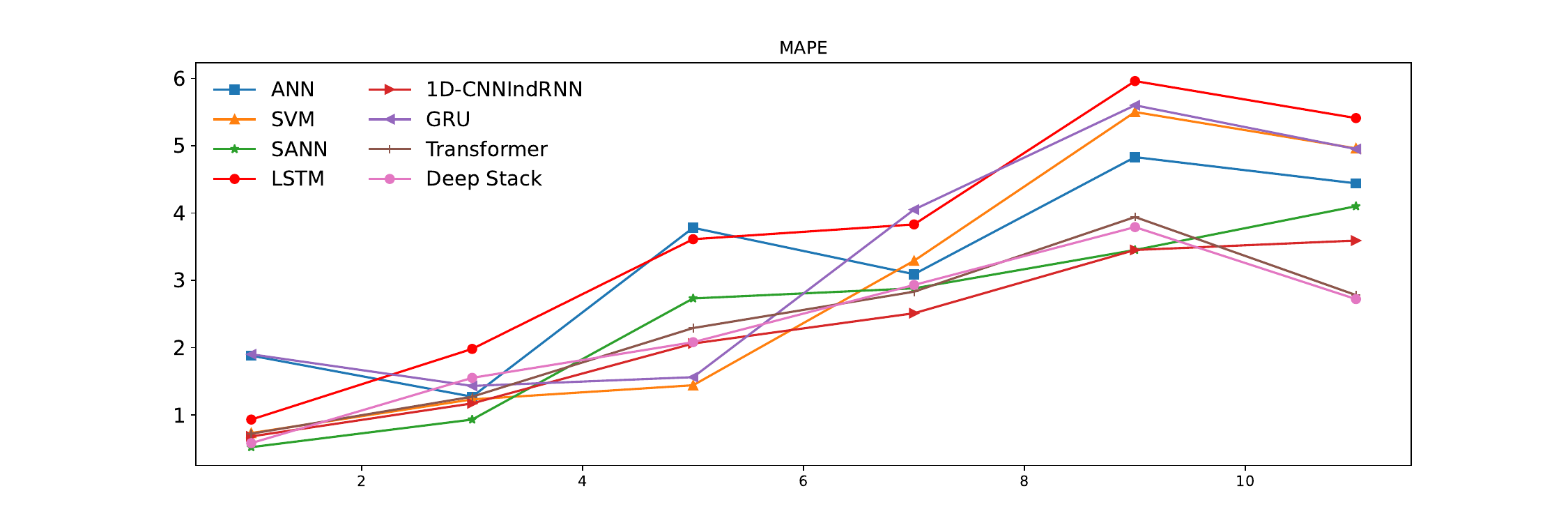}}
\caption{MAE, RMSE, and MAPE of the price prediction of the proposed models and baseline for forecast in Interval  I, II, and III.(note: $x[1]=I$,$x[3]=II$,$x[5]=III$,$x[7]=7-days$,$x[9]=30-days$,and $x[11]=90-days$) }
\label{F6dd}
\end{figure*}

Also, Figure \ref{F6d} shows the values of MAE, RMSE, and MAPE for feature selection models. In these approaches, the Embedded approach provided weaker results. The worst results obtained are related to the 30th interval of the Embedded approach. Also, the comparative results of the deep stack approach and other baseline approaches are shown in Figure \ref{F6dd}. The worst result in the regression problem is related to the LSTM approach.

\section{Conclusion}

BTC price modelling is done in two general ways: 1)price increase and decrease, which is a classification problem(binary classification), and 2) real price, which is a regression problem. This paper presents a deep stack model that can solve price forecasting problems with a very low error rate. However, the classification problem that was a 2-class classification is still an open challenge for all researchers. Since the price of Bitcoin is random and no particular set of features can provide a perfect prediction, researchers have been forced to use different methods of feature selection, feature extraction, and feature creation. In this article, we tried to use the features related to blockchain. Technical indicators(TI) are simple mathematical tools for converting these raw features into time series features that can be used for fundamental estimations.. After applying technical indicators on input features, Chi2, RFE, and Embedded feature selection approaches were applied. The data was collected in three intervals and three multi-day forecasts. The combination of the proposed stack-based approach and chi2 obtained more acceptable results than the existing approaches in the literature. The deep stack model also provides a useful view of classification and can predict increases and decreases with an accuracy of over 80\%.

One of the most obvious research in this field is in the classification section, which uses classes 1 and 0 for increase and decrease. This means that if the next price has increased by even \$0.01 in the real space, it will force the trader to enter the trade. As one of the future works, it can use the range of properties of changes in the classification, which also takes into account the sensitivity of entering the trade. Another problem of time series classification is unbalanced class. The challenge in these issues is that some intervals or classes rarely occur. This problem is an unbalanced class problem for which the following methods can be used:

\begin{itemize}
\item \textbf{Resampling: } Various algorithms have been proposed for this purpose. For example, algorithms \cite{Sa1}\cite{Sa2}\cite{Sa3}\cite{Sa4}, and \cite{Sa5} can be used.
  \item \textbf{High-dimensional Imbalanced Time-series classification (OHIT)}\cite{Sa6}:
OHIT first uses a density ratio-based joint nearest neighbor clustering algorithm to capture minority class states in a high-dimensional space. Depending on different clustering algorithms, this clustering can get different results. It then for each mode applies the shrinkage technique of large-dimensional covariance matrix to obtain an accurate and reliable covariance structure. Finally, OHIT generates structure-preserving synthetic samples based on a multivariate Gaussian distribution using the estimated covariance matrices.
  \item \textbf{IB-GAN}\cite{Sa7}: IB-GAN uses imputation and resampling techniques to generate higher quality samples from randomly masked vectors than white noise, and balances the classifier through a pool of real and synthetic samples. Hyperparameter imputation pmiss allows to regularize the classifier variation by adjusting the innovations introduced through generator imputation. IB-GAN is simple to train and model, pairing each deep learning classifier with a generator-discriminator pair, resulting in higher accuracy for less observed classes. The basis of this approach is a GAN network that tries to generate cases similar to the minority class.
\end{itemize}

\end{document}